\documentclass{article}

\usepackage{PRIMEarxiv}

\usepackage{array}

\usepackage[utf8]{inputenc} % allow utf-8 input
\usepackage[T1]{fontenc}    % use 8-bit T1 fonts
\usepackage{hyperref}       % hyperlinks
\usepackage{url}            % simple URL typesetting
\usepackage{booktabs}       % professional-quality tables
\usepackage{amsfonts}       % blackboard math symbols
\usepackage{nicefrac}       % compact symbols for 1/2, etc.
\usepackage{microtype}      % microtypography
\usepackage{lipsum}
\usepackage{fancyhdr}       % header
\usepackage{graphicx}       % graphics
\graphicspath{{media/}}     % organize your images and other figures under media/ folder
\usepackage{algorithm} % added by fujii
\usepackage{algpseudocode}
\usepackage{color}% added by fujii
\usepackage{arydshln} % added by fujii
\usepackage{amsfonts} % added by fujii
\usepackage{abstract}
\usepackage{amsmath}
\usepackage{multirow}
\usepackage{balance}
\usepackage{authblk} % ← これが鍵！
\usepackage{placeins}

\usepackage{titlesec}
\usepackage{multicol}
\usepackage{float}

\usepackage{authblk}
\usepackage{abstract}

\makeatother

\title{A Straightforward Gradient-Based Approach for High-Tc Superconductor Design:\\ Leveraging Domain Knowledge via Adaptive Constraints}

\author[1]{Akihiro Fujii$^*$}
\author[1,2]{Anh Khoa Augustin Lu}
\author[3]{Koji Shimizu}
\author[1]{Satoshi Watanabe}

\affil[1]{Department of Materials Engineering, The University of Tokyo}
\affil[2]{National Institute for Materials Science (NIMS)}
\affil[3]{National Institute of Advanced Industrial Science and Technology (AIST)}

\date{} % 日付を表示したくなければ空に

\begin{document}

\twocolumn[
  \begin{@twocolumnfalse}
  \maketitle
  \vspace{-5.0em}  % ★ここがポイント
\begin{center}

    \textit{* Corresponding author: akihiro.fujii@cello.t.u-tokyo.ac.jp}
    \end{center}
    
    \begin{abstract}
    \vspace{1em} % ← Abstract と次の要素（keywordsや本文）との間隔
    \noindent
    Materials design aims to discover novel compounds with desired properties. However, prevailing strategies face critical trade-offs. Conventional element-substitution approaches readily and adaptively incorporate various domain knowledge but remain confined to a narrow search space. In contrast, deep generative models efficiently explore vast compositional landscapes, yet they struggle to flexibly integrate domain knowledge. To address these trade-offs, we propose a gradient-based material design framework that combines these strengths, offering both efficiency and adaptability. In our method, chemical compositions are optimised to achieve target properties by using property prediction models and their gradients. In order to seamlessly enforce diverse constraints—including those reflecting domain insights such as oxidation states, discretised compositional ratios, types of elements, and their abundance, we apply masks and employ a special loss function, namely the integer loss. Furthermore, we initialise the optimisation using promising candidates from existing dataset, effectively guiding the search away from unfavourable regions and thus helping to avoid poor solutions. Our approach demonstrates a more efficient exploration of superconductor candidates, uncovering candidate materials with higher critical temperature than conventional element-substitution and generative models. Importantly, it could propose new compositions beyond those found in existing databases, including new hydride superconductors absent from the training dataset but which share compositional similarities with materials found in literature. This synergy of domain knowledge and machine-learning-based scalability provides a robust foundation for rapid, adaptive, and comprehensive materials design for superconductors and beyond.
    \end{abstract}

    \vspace{2em} % optional spacing after abstract
  \end{@twocolumnfalse}
]


%%%MAIN TEXT%%%%
\section{Introduction}

\begin{figure*}[t]
\centering
  \includegraphics[width=0.80\textwidth]{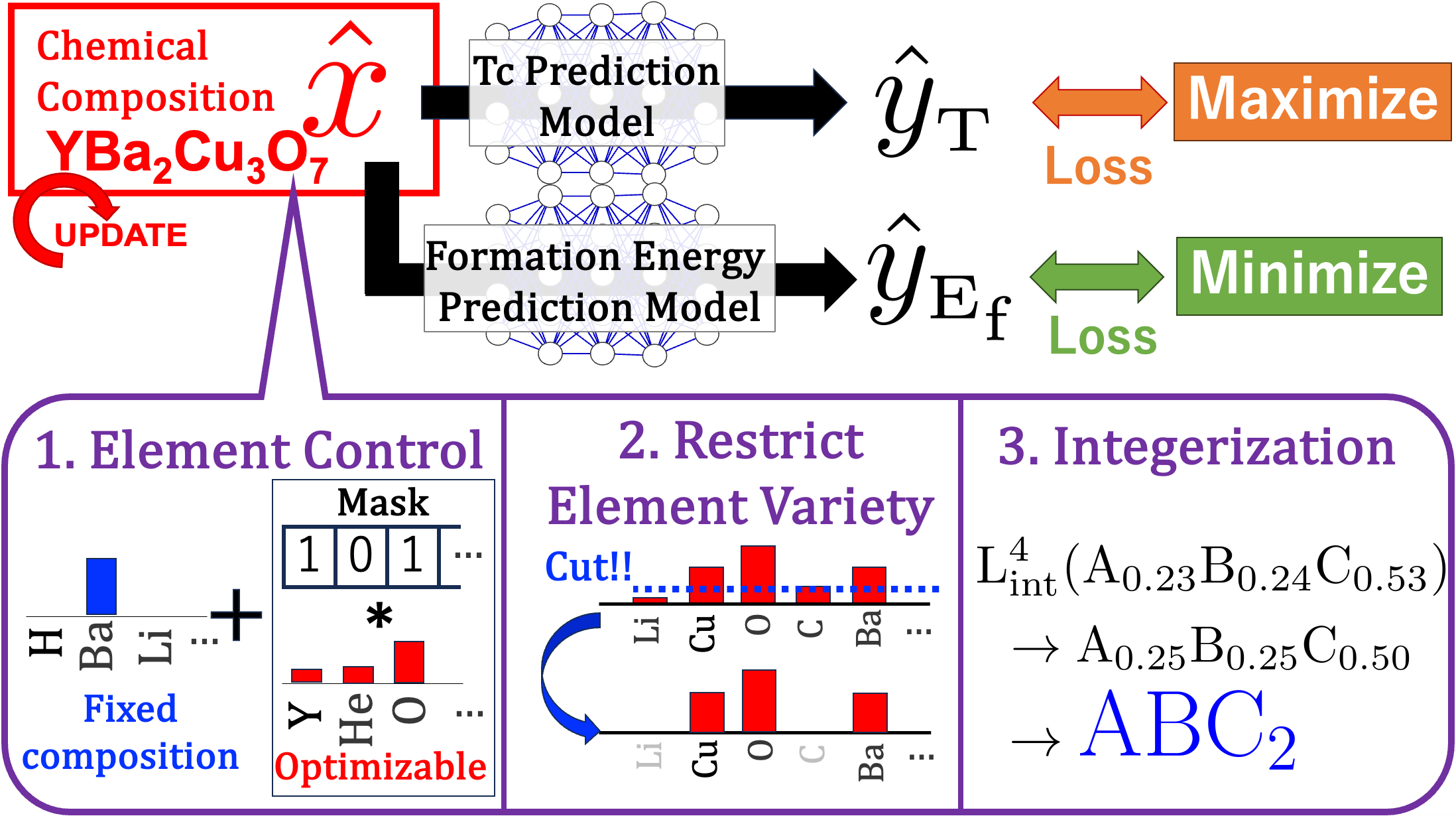}
  \caption{Overview of Knowledge-Integrated Adaptive Gradient-based Optimisation (KIAGO). KIAGO simultaneously maximises $T_c$ and minimises the formation energy by optimising the input composition using two pretrained models and their gradients. Through the use of fixed composition vectors, masks, and specialised loss functions, KIAGO enables flexible control of the composition in three ways: \textbf{1. Element control.} Specific elements can be fixed and excluded from the optimisation target to perform conditional optimisation. KIAGO is also able to control which elements appear during the optimisation via masks. Here, we fix the composition of Barium and exclude Helium from the optimisation using the mask; \textbf{2 Restricting the maximum number of elements.} We first rank elements by their abundance in the composition and create a mask to keep only the most abundant ones up to a specified cutoff. All other elements are set to zero, ensuring that the total number of elements never exceeds the chosen limit. In this figure, we select the three most abundant elements to build a mask, which then restricts the final composition to those three elements. \textbf{3. Normalising the compositional ratios to small integers.} Here, we use the loss function $L_{int}^4$ to guide the normalised composition to a composition consisting of four atoms.}
  \label{figure__overview}
\end{figure*}

Materials design is crucial for technological innovation such as the discovery of new superconductor materials. High-temperature superconductors (HTS) are especially promising because they reduce cooling costs and enable higher magnetic fields. They are also expected to be applied in fusion power generation, electric power cables, and superconducting maglev trains~\cite{hull2003applications,uglietti2019review}. 

Exploiting physical insights—such as selection of elements based on their oxidation states and the fact that materials with too many elements are impractical—can narrow this search, making materials design more efficient. A traditional technique in materials design is elemental substitution (i.e., doping)~\cite{skakle1998crystal,yao2004doping,erwin2005doping,terzioglu2019investigation}. In this approach, one starts with a promising host material and partially substitutes certain elements to tune the properties. Substituted elements are typically chosen based on physical insights—such as oxidation states—to ensure charge neutrality and other key constraints. 

Machine learning (ML) has become a widely used approach for materials discovery, offering faster property predictions than conventional Density Functional Theory (DFT) calculations and thus enabling high-throughput screening. In the context of HTS development, some studies~\cite{stanev2018machine,matsumoto2019acceleration,zeng2019atom,dan2020computational,le2020critical,konno2021deep,zhang2022machine,taheri2022prediction,zhang2023integrated} have focused on training superconducting transition temperature ($T_c$) prediction models using the SuperCon dataset~\cite{supercon}, which comprises a large set of known superconductors.

Recently, deep generative models have gained prominence in materials design~\cite{xie2021crystal,ren2022invertible,zeni2023mattergen,yang2024novo}, including the quest to discover novel superconductors~\cite{zhong2023deep,kim2023scgan}. These models propose new compounds by learning the statistical distribution of existing data, thus enabling the exploration of a vast chemical space. Several studies~\cite{wines2023inverse,yuan2024diffusion} employ diffusion models~\cite{ho2020denoising}—a deep generative model widely used in the computer vision field~\cite{croitoru2023diffusion}—to generate superconductor candidates. SuperDiff~\cite{yuan2024diffusion}, a diffusion model for superconductors, generates candidate superconductors by gradually removing noise from a noisy composition. Moreover, SuperDiff can generate conditioned outputs based on reference compounds using Iterative Latent Variable Refinement (ILVR)~\cite{choi2021ilvr}. While conventional generative methods only explore materials within existing databases, SuperDiff can generate new materials based on promising reference compounds.

Moreover, there are strategies that guide deep generative models toward desired properties, such as label-based conditional generation~\cite{brock2018large}, Universal Guidance~\cite{Bansal_2023_CVPR} (UG), Classifier Guidance (CG)~\cite{dhariwal2021diffusion} and Classifier-Free Guidance (CFG)~\cite{ho2022classifier}. While CG and UG use a separate property predictor to steer the generation process, CFG does not require such a predictor. Although the label-based conditional generation and CFG have both been extensively validated in image generation, their reliance on labels within the dataset limits their flexibility in materials design. By contrast, CG can be conditioned on labels not present in the target dataset using models trained on other datasets. Xie \textit{et al.}~\cite{xie2021crystal} employ a strategy similar to CG, combining a diffusion model with a formation energy prediction model. Applying CG to $T_c$ prediction models and superconducting material generation models such as SuperDiff has the potential to enable HTS design.

A gradient-based method~\cite{ren2020benchmarking,allen2022physical,hwang2022solving,fujii2023enhancing} that uses prediction models and their gradients to optimise inputs has recently attracted attention. This method is similar to CG and UG but simpler, as it does not require training a generative model. Moreover, this method allows for more flexible and adaptive conditional optimisation~\cite{fujii2023enhancing}. While there is no study of applying this technique to composition optimisation, it could be a promising approach for materials design.

Despite these advances, significant trade-offs remain. Elemental substitution can incorporate physical knowledge but may limit exploration to a relatively narrow search space. Deep generative models can explore a broader chemical space efficiently, yet they struggle to flexibly integrate physical knowledge—such as atomic valence constraints or converting compositional ratios to integers—in an adaptive manner. On the other hand, the gradient-based method has a risk of falling into poor solutions, though this method has the potential to introduce various physical knowledge in an adaptive manner. 

In this paper, to address these issues, we adopt a gradient-based method and propose a straightforward materials-design method called Knowledge-Integrated Adaptive Gradient-based Optimisation (KIAGO). This framework combines the adaptive application of domain knowledge with computational efficiency to directly optimise chemical compositions (Fig.~\ref{figure__overview}). KIAGO does not require training a deep generative model, making it more straightforward to implement. Specifically, we adopt two property prediction models—one for $T_c$ and another for formation energy-to maximise $T_c$ while enhancing stability, thereby proposing realistic materials. Unlike CFG and label-based conditional generation, KIAGO can optimise formation energy (which is not included in the SuperCon dataset) by using a separately trained formation energy prediction model. Additionally, by conducting an intensive search around promising materials, KIAGO mitigates the risk of being trapped at poor results. Moreover, by applying masks and a specialised loss function to enforce integer values, we can effectively embed physical insights (e.g., ensuring the retention of specific elements, oxidation states, the number of elements, or integer compositional ratios), thus providing a versatile framework that accommodates diverse constraints in adaptive manners.

To validate the effectiveness of KIAGO, we performed experiments to propose promising HTS. Our approach significantly outperformed both generative models (SuperDiff and SuperDiff with CG) and conventional elemental substitution techniques in proposing high-$T_c$ candidates efficiently. In particular, we found that SuperDiff with CG tended to generate materials with lower $T_c$ values because the $T_c$ distribution in the original data constrained them. In contrast, our method proposed high-$T_c$ candidates without being limited by the original distribution. Additionally, KIAGO could keep some part of the composition fixed while optimising others, relevantly replace elements according to their oxidation states, and maintain charge neutrality perfectly. Additionally, KIAGO proposed candidate compositions that shared the same elements as hydride superconductors reported in other literature despite their absence from the SuperCon dataset. These results highlight its potential for discovering novel materials.

\section{KIAGO}

\subsection{Overview}
Knowledge-Integrated Adaptive Gradient-Based Optimisation (KIAGO) is a gradient-based method that uses pre-trained models and their gradients to directly optimise the input representation—in this case, the normalised compositional vector of candidate materials. Rather than merely searching for compositions that yield favourable properties,  KIAGO introduces three key strategies to enhance material quality and provide fine-grained control: 1. Initialisation from promising materials to mitigate the risk of being trapped at poor results; 2. Masking to control elemental types; 3. Special loss functions for conversion to integers and atomic-count constraints. 

\subsection{Gradient-based method}
A gradient-based method can adopt any predictive model, provided the chain rule of differentiation is valid from input to output. To propose superconducting materials with high-$T_c$, we employ a $T_c$ predictor $f_{T_c}$. We also use a formation-energy predictor $f_{E_f}$ to propose compounds that are both high-$T_c$ and thermodynamically feasible. We introduce a hyperparameter $\alpha$ and define the loss $L$ as
\begin{gather}
L = -f_{T_c}(\hat{x})+ \alpha f_{E_f}(\hat{x}) \label{eq___naive_loss} \\ 
\hat{x}_* = \underset{\hat{x}} {\operatorname{argmin}} \ L.
\end{gather}
Here, $\hat{x} \in [0,1]^{N_{\mathrm{elem}}}$ is a compositional vector spanning $N_{\mathrm{elem}}$ elements. Minimising $L$ aims to increase $T_c$ while lowering the formation energy. However, simple minimisation poses several issues: (1) it may converge to poor solutions, (2) it lacks control over the number and type of elements, and (3) it does not ensure integer ratios in the final composition.

\subsection{Initialisation based on promising materials} \label{section__initialisation} 
To avoid converging to poor solutions, we adopt a strategy of starting the optimisation from various initial states including those corresponding to known promising materials. We can reduce this risk by focusing on the areas of existing high-performance compounds. Such a strategy goes beyond doping-like approaches that only alter part of an existing material, enabling a broader range of materials to be explored. Specifically, we perturb known superconductors by substituting elements randomly and adding new elements to the composition. This technique effectively explores the local neighbourhood of promising materials.

\subsection{Controlling the types of elements present} \label{section__mask}
We next control which elements appear in the composition by combining a fixed composition vector and a mask (Fig.~\ref{figure__overview}-1). First, we split $\hat{x}$ into a fixed portion $x_\mathrm{const}$ and an optimisable portion $\hat{x}_\mathrm{opt}$:
\begin{gather}
\hat{x}_\mathrm{opt} \in \mathbb{R}^{N_{\mathrm{elem}}}, \ x_\mathrm{const} \in [0,1]^{N_{\mathrm{elem}}}, \ \sum_i x^i_{\mathrm{const}} < 1 \\ \hat{x} = x_\mathrm{const} + \sigma(\hat{x}_\mathrm{opt}) (1-\sum_i x^i_{\mathrm{const}}).
\end{gather}
Here, $\sigma$ is a normalisation function that ensures each element is non-negative and the total sum is 1. A possible approach was to use the softmax function. However, to emphasize elements that remain unused, we instead chose to use normalisation after applying Rectified Linear Unit (ReLU).
\begin{gather}
\sigma(x) = \frac{\mathrm{ReLU}(x)}{\sum\mathrm{ReLU}(x)}
\end{gather}
Because $x_\mathrm{const}$ remains unchanged, its specified composition remains fixed during optimisation. We further introduce a mask $M_{\mathrm{elem}} (M_{\mathrm{elem}} \in \{0,1\}^{N_{\mathrm{elem}}})$ to select the allowable elements. Concretely, we set 
\begin{gather}
\hat{x}_\mathrm{opt} = \hat{x}_\mathrm{base} *M_\mathrm{elem},
\end{gather}
where $\hat{x}_\mathrm{base} (\hat{x}_\mathrm{base} \in \mathbb{R}^{N_{\mathrm{elem}}})$ is a trainable parameter, and the asterisk ($*$) denotes element-wise multiplication. This mask enforces strict control over which elements can be used, thus guiding the optimisation toward compositions that meet specified domain constraints.

\subsection{Controlling the number of element types in the composition} \label{section__max_elem}
 
To achieve realistic composition, we use a mask to limit how many elements can appear in the composition (Fig.~\ref{figure__overview}-2). To do this, we sort the compositional values in ascending order and create a mask $M^{\mathrm{max}}_{\mathrm{elem}} \ (M^{\mathrm{max}}_{\mathrm{elem}} \in \{0,1\}^{N_{\mathrm{elem}}})$, which sets to zero any element index beyond the allowed maximum elements $\mathrm{n^{max}_{elem}}$. 
\begin{gather}
\sum_{i,j} (M^{\mathrm{max}}_{\mathrm{elem}})_{i,j} = \mathrm{n^{max}_{elem}} \\
\hat{x}'_\mathrm{opt} = \hat{x}_\mathrm{opt} * M^\mathrm{max}_\mathrm{elem}, 
\end{gather}
\subsection{Integer loss} 

We construct a loss function that guides the composition into integer-compatible values during optimisation (Fig.~\ref{figure__overview}-3). Such values $\lbrace c^{N_{\mathrm{unit}}}_n\rbrace$ are those for which the product of the normalised composition and the unit cell size $N_{\mathrm{unit}}$ becomes an integer ratio. For instance, if $N_{\mathrm{unit}}=4$, then the feasible set $\lbrace c^4_n\rbrace$ is $\lbrace 0.00, 0.25, 0.50,0.75, 1.00 \rbrace$.

\begin{figure}[t]
\centering
  \includegraphics[width=0.45\textwidth]{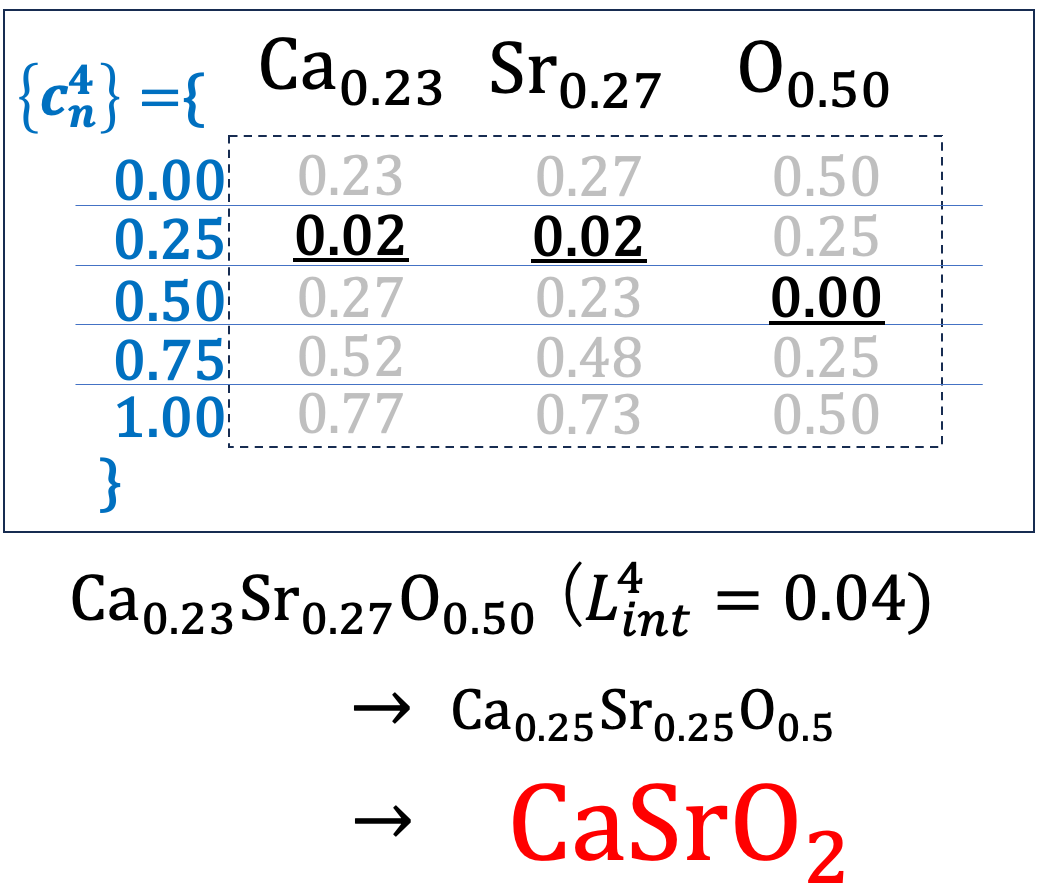}
  \caption{ An overview of the integer loss $L_{int}^{4}$ under the assumption that each unit cell contains four atoms. The numbers shown inside the dashed box represent all possible difference combinations between the integer-compatible set $\lbrace c^4_n\rbrace$ and the composition values. The total loss is obtained by selecting the minimum among these combinations for each element (indicated by the black underline) and summing them.}
  \label{figure__integer_loss}
\end{figure}

The integer loss measures how far each compositional ratio of element $i$ ($\hat{x}^i$) is from its nearest value in $\lbrace c^{N_{\mathrm{unit}}}_n\rbrace$. 
\begin{gather}
\lbrace c^{N_{\mathrm{unit}}}_n \rbrace = \lbrace n/N_{\mathrm{unit}} \rbrace_{n=0,1,...,N_{\mathrm{unit}}} \\  \hat{x}^i \in \lbrace \hat{x}^{\mathrm{H}},\hat{x}^{\mathrm{He}},\hat{x}^{\mathrm{Li}},...,\hat{x}^{N_{\mathrm{elem}}} \rbrace \\ L_{int}^{N_{\mathrm{unit}}} (\hat{x})= \sum_{i=1}^{N_{\mathrm{elem}}} \underset{n}{\mathrm{min}}\Bigr|\hat{x}^i-c^N_n\Bigl|
\end{gather}
As an example, Fig.~\ref{figure__integer_loss} shows the result of applying $L_{int}^{4}$ to $\mathrm{Ca_{0.23}Sr_{0.27}O_{0.50}}$. We assume $N_{\mathrm{unit}}=4$ and guide the compositions toward the nearest values in $\lbrace 0.00, 0.25, 0.5,0.75, 1.00 \rbrace$.
Because it is difficult to fix $N_{\mathrm{unit}}$ in advance, we evaluate multiple candidates for $N_{\mathrm{unit}}$ and choose the one that yields the smallest loss. Concretely, we define $L_{integer}$ as follows:
\begin{gather}
L_{\mathrm{integer},\lbrace N_\mathrm{unit}\rbrace} (\hat{x})= \underset{N \in \lbrace N_\mathrm{unit}\rbrace}{\mathrm{min}} \ L_{int}^N (\hat{x}). \label{eq__integer_loss}
\end{gather}
This flexible approach selects a suitable integer grid even when the optimal cell size is unknown.

\subsection{Optimisation procedure}

KIAGO divides its optimisation into two stages. First, as in Section~\ref{section__initialisation} and \ref{section__mask}, we construct an initial $\hat{x}_\mathrm{base}$ and control which elements appear while iteratively minimising the following loss $L_{\mathrm{1st}}$. This process yields $\hat{x}^{\mathrm{1st}}_*$.
\begin{gather}
\hat{x} = x_\mathrm{const} + \sigma( \hat{x}_\mathrm{base} *M_\mathrm{elem}) (1-\sum_i x^i_{\mathrm{const}})  \label{eq__first_sgate_srt} \\
L_{\mathrm{1st}} = -f_{T_c}(\hat{x})+\alpha f_{E_f}(\hat{x}) \label{loss_for_superdiff_cg}\\
\hat{x}^{\mathrm{1st}}_* = \underset{\hat{x}_\mathrm{base}} {\operatorname{argmin}} \ L_{\mathrm{1st}}  \label{eq__first_sgate_end}
\end{gather}
Next, we introduce the conversion to integers and a maximum-atom constraint. We use $\hat{x}^{\mathrm{1st}}_*$ to build the mask $M^\mathrm{max}_\mathrm{elem}$, then iteratively minimise the loss $L_{\mathrm{2nd}}$. 
\begin{multline}
\hat{x}'(\hat{x}_\mathrm{base}) = x_\mathrm{const} + \\  \sigma( \hat{x}_\mathrm{base} *M^\mathrm{max}_\mathrm{elem}*M_\mathrm{elem}) (1-\sum_i x^i_{\mathrm{const}}) \label{eq__second_sgate_srt}
\end{multline}
\begin{gather}
L_{\mathrm{2nd}} = -f_{T_c}(\hat{x}')+ \alpha f_{E_f}(\hat{x}') + \beta ÷
L_{integer,\lbrace N_\mathrm{unit}\rbrace}(\hat{x}') \label{eq__second_sgate_loss} \\   
\hat{x}^{\mathrm{2nd}}_* = \underset{\hat{x}_\mathrm{base}} {\operatorname{argmin}} \ L_{\mathrm{2nd}} \label{eq__second_sgate_end}
\end{gather}
Here, $\beta$ is a hyperparameter. We take $\hat{x}'(\hat{x}^{\mathrm{2nd}}_*)$ as the final solution.

\section{Results}

\subsection{Implementation Details} \label{section___implementatin_details}
We used PyTorch~\cite{paszke2017automatic} to implement KIAGO. KIAGO optimises a total of 4,096 candidate compositions across 1,000 steps using Adam optimiser~\cite{kingma2014adam}. We first perform 500 steps of optimisation using equtations~\ref{eq__first_sgate_srt} to \ref{eq__first_sgate_end}, followed by another 500 steps using equations.~\ref{eq__second_sgate_srt} to \ref{eq__second_sgate_end} with $\alpha=4$ (selected based on tuning) and $\beta=1$. To predict the superconducting transition temperature ($T_c$), we employ a ResNet18 model~\cite{he2016deep} trained on normalised compositions of SuperCon and Crystallography Open Database (COD)~\cite{vaitkus2023workflow,merkys2023graph,vaitkus2021validation,quiros2018using,merkys2016cod,gravzulis2015computing,gravzulis2012crystallography,gravzulis2009crystallography,downs2003american}. Each composition is represented by a periodic table-based feature map, which has four channels corresponding to the s, p, d, and f orbitals~\cite{konno2021deep}. For each element, we set flags at the positions of its row and column on the periodic table, as well as at its relevant orbital channels. We then multiply these element-level feature maps by the respective compositional ratios to create the final input representation. Further details are available in Section~\ref{section__tc_prediction_model}. For formation energy, we used ElemNet~\cite{jha2018elemnet}, which is originally implemented in Tensorflow 1.x~\cite{tensorflow2015-whitepaper} and we re-implemented it in PyTorch. Since ElemNet only covers elements up to atomic number 86, we apply a mask to exclude elements beyond that range. Additional technical specifics are given in Section~\ref{section__elemnet}.
% [修正MEMO] ElemNetは先行研究で利用されていることを言及し、SuperDiffでは（確か）0.0 eV以下でやっていたので、それに合わせたことを言及する。Pytorchを使ったことと、そのバージョンに言及する

We compared KIAGO against two baselines: a conventional elemental-substitution (C-ES) approach and SuperDiff, a diffusion-based generative model. The C-ES method randomly replaces some elements with others of identical oxidation states. For SuperDiff, we used the official implementation and train on the same data for our $T_c$ prediction model, but without normalising compositions. Following the official code, we conducted 1,000 diffusion steps. According to the original SuperDiff, we conditioned generations of compositions on existing superconductors using Iterative Latent Variable Refinement (ILVR). We applied scale factors of 1, 2, 4, and 6 to yield a total of 4,096 samples. We also incorporate Classifier Guidance (CG) using $T_c$ prediction model and ElemNet into SuperDiff to compare it directly with KIAGO. Normally, the classifier used for CG must be trained on data with noise, which would make Universal Guidance (UG) the better choice for off-the-shelf models. However, we found that UG did not work well and CG still improved $T_c$ using models without noise-augmented training. Therefore, we decided to use CG. Although neither model is strictly a classifier, we refer to this approach as CG for convenience. Note that this is not proposed in the original paper\cite{yuan2024diffusion}. At each inference step, we used equation~\ref{loss_for_superdiff_cg} with $\alpha=4$ for gradient guidance, and we tune the guidance weight from 1\texttimes10\textsuperscript{-7} to 1.0, ultimately selecting 1\texttimes10\textsuperscript{-3}. Additional details of SuperDiff are provided in Section \ref{section__apx__superdiff}.
% -> Additional details of SuperDiff are provided in \ref{section__apx__superdiff}.

After generating candidate compositions, we applied a multi-step screening procedure to ensure realistic materials. First, we used SMACT~\cite{davies2019smact} to filter compositions with charge neutrality and electronegativity balance, following Yuan \textit{et al.}~\cite{yuan2024diffusion}. Next, we selected only those with formation energies (predicted by ElemNet) less than zero. We also removed compositions containing ten or more elements since the preprocessed SuperCon data have at most nine. Finally, we evaluate $T_c$ values using the same ResNet18 predictor used in KIAGO and SuperDiff with CG.

\subsection{Mitigating the risk of convergence to non-promising solutions} \label{section__test_of_init}

In this section, we investigated whether our initialisation scheme could mitigate the risk of converging to non-promising local minima in the first stage (equations.~\ref{eq__first_sgate_srt} to \ref{eq__first_sgate_end}). Specifically, we aimed to determine whether our method can produce more promising local optima than a purely random initialisation. In our method, we began with the superconductor $\mathrm{CuLa_2O_4}$ from the SuperCon dataset. With a probability of 0.22, we replaced elements of its composition with different elements chosen according to their occurrence frequencies in SuperCon. We then selected random elements with random compositional ratios (from 0.0 to 0.3) for those elements, normalised the resulting composition, and used it as the initialisation. By contrast, the random initialisation selects seven elements, to match the number of atoms in $\mathrm{CuLa_2O_4}$, uniformly at random and assigns them random compositional values.

\begin{figure}[t]
\centering
  \includegraphics[width=0.45\textwidth]{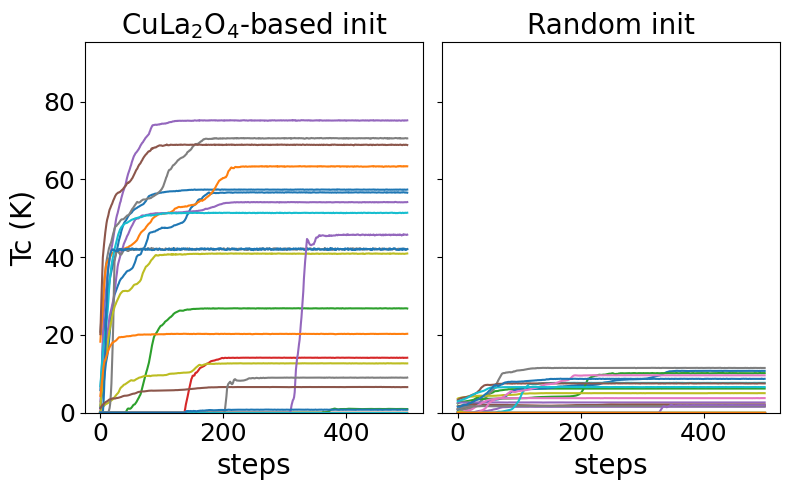}
  \caption{Comparison of optimisation results under different initialisation methods. Both approaches employ Adam optimiser~\cite{kingma2014adam} with a learning rate of 0.001. (Left) Initialisation by adding noise to an existing superconductor ($\mathrm{CuLa_2O_4}$). (Right) Random initialisation, in which seven elements are chosen arbitrarily and assigned random compositional values.}
  \label{figure__initialisation_comparison}
\end{figure}

Fig.~\ref{figure__initialisation_comparison} shows the optimisation results. Our initialisation scheme yields higher $T_c$ values than random initialisation. Although our method can still become trapped in local optima, it proposes more promising solutions than the random approach. Hence, our approach partially mitigates the inherent challenge of local minima in gradient-based methods.

\subsection{Converting compositional ratios to integers via loss-based approach}

In this section, we compared integer conversion method based on a loss function $L_{integer,\lbrace N_\mathrm{unit}\rbrace}$ with a rule-based integer conversion method. We aimed to determine which approach reduced the drop in $T_c$ in the second stage (equations.~\ref{eq__second_sgate_srt} to \ref{eq__second_sgate_end}). In the loss-based method, we follow equation~\ref{eq__second_sgate_loss} to maximise $T_c$ while minimising $E_f$. Specifically, we guided the composition toward an integer representation by selecting an optimal total number of atoms from a set of integers smaller than the specified maximum atom count. We optimised them for 500 steps for integer conversion. By contrast, the rule-based approach multiplies the normalised composition by the specified number of atoms and then rounds each value to the nearest integer.

Table~\ref{table__effectiveness_of_integer_loss} shows the results comparing the rule-based approach and $L_{integer,\lbrace N_\mathrm{unit}\rbrace}$.  Because it remains closer to the pre-conversion to integers composition, rounding with a larger total number of atoms is generally advantageous. Although the rule-based method fully exploits this by always rounding at the maximum atom count, $L_{integer,\lbrace N_\mathrm{unit}\rbrace}$ does not always do so, yet it still often performs better. 

\begin{table}
\centering \small
\caption{Comparison between the rule-based approach and $L_{integer,\lbrace N_\mathrm{unit}\rbrace}$ for converting compositional ratios to integers. The table shows the average change in $T_c$ before and after conversion to integers under certain maximum numbers of atoms, based on a total of 61,440 samples derived from 15 different superconducting materials.}
\label{table__effectiveness_of_integer_loss}
\begin{tabular}{l|cc}
\hline
    Max Num Atoms & $L_{integer,\lbrace N_\mathrm{unit}\rbrace}$ (K) & Rule-based (K) \\
    \hline
    15 & \underline{\textbf{-6.44}} & $-$8.22 \\ 
    20 & \underline{\textbf{-2.47}} & $-$3.27 \\ 
    25 & \underline{\textbf{-2.36}} & $-$3.40 \\ 
    50 & $-$3.22 & \underline{\textbf{-1.13}} \\ 
    100 & $-$0.88 & \underline{\textbf{-0.27}} \\ 
    \hline
\end{tabular}
\end{table}

\subsection{Generating superconductors with higher $T_c$ based on existing ones} \label{section__based_on_existing}

In this section, we investigated whether our method could propose superconductors with higher $T_c$ values based on known superconductors as initial candidates. For KIAGO, we start with existing superconductors and introduce noise to the compositions described in Section~\ref{section__test_of_init}. We used Adam optimiser with learning rate of 0.03 for KIAGO. We set $\{ N_\mathrm{unit}\}=\{ 1,2,...,25\}$ and $ \mathrm{n^{max}_{elem}}=25$. SuperDiff conditions on existing superconductors via Iterative Latent Variable Refinement (ILVR). The conventional elemental-substitution (C-ES) approach randomly replaces a subset of elements with others sharing the same oxidation state. For each base material of superconductor, we select copper-based, iron-based, and other superconductors that pass charge-neutrality and electronegativity screening by SMACT, then randomly choose from these sets as base materials.

Table~\ref{table__result_of_exsiting_composition_based} presents the differences in predicted $T_c$ between the generated superconductors and their base materials. KIAGO achieves the most efficient exploration of higher $T_c$ values compared to other methods. By contrast, there are experiments where SuperDiff does not yields any valid materials passing all screenings. This limitation may stem from the fact that many entries in SuperCon do not pass charge-neutrality and electronegativity checks; hence, the model struggles to generate valid compositions. The C-ES method also fails to propose sufficiently high $T_c$ compounds, likely because its rule-based approach cannot fully explore the vast compositional space. In contrast, KIAGO proposes many materials showing substantial $T_c$ increases. For completeness, Section~\ref{section__detailed_results} includes all screening-pass rates. 

Table~\ref{table__result_of_samples_of_exsiting_composition_based} describes example compositions. Both KIAGO and C-ES yield integer-total compositions, allowing straightforward induction of possible crystal structures. However, SuperDiff and SuperDiff with CG frequently produce non-integer totals, making immediate structural analysis more challenging.

\begin{table}
\centering \small
\caption{Differences in $T_c$ (K) between the average $T_c$ of the highest Top-30 proposed superconductors and the base superconductors in the experiments of proposing superconductors based on existing ones. SD, SD w/ CG, and C-ES denote SuperDiff, SuperDiff with Classifier Guidance, and conventional elemental substitution, respectively. After screening 4096 samples, the $T_c$ prediction model was used to calculate $T_c$. 'N/A' indicates that none of the samples passed the screening.}
\label{table__result_of_exsiting_composition_based}
\resizebox{0.5\textwidth}{!}{
\begin{tabular}{l|wc{1cm}wc{1cm}wc{1cm}wc{1cm}}
\hline
    \begin{tabular}{c}Base materials \\ from SuperCon\end{tabular} & \begin{tabular}{c}KIAGO\\Top-30\\ $\Delta T_c$(K) \end{tabular} & \begin{tabular}{c}SD\\Top-30\\ $\Delta T_c$(K) \end{tabular} & \begin{tabular}{c}SD w/ CG\\Top-30\\ $\Delta T_c$(K) \end{tabular} & \begin{tabular}{c}C-ES\\Top-30\\ $\Delta T_c$(K) \end{tabular} \\
    \hline
    $\mathrm{LaNiAsO}$ & \underline{\textbf{97.64}} & $-$1.42 & $-$0.48 & 16.79 \\ 
    $\mathrm{SrFe_{1.88}Ni_{0.12}As_{2}}$ & \underline{\textbf{86.77}} & 26.76 & 0.74 & 18.96 \\ 
    $\mathrm{Sr_{4}V_{2}Fe_{2}As_{2}O_{6}}$ & \underline{\textbf{89.89}} & $-$13.77 & $-$3.99 & 0.00 \\ 
    $\mathrm{LaPt_{2}B_{2}C}$ & \underline{\textbf{86.39}} & $-$5.01 & $-$5.27 & 5.06 \\ 
    $\mathrm{HgBa_{2}Ca_{2}Cu_{3}O_{8}}$ & \underline{\textbf{16.70}} & $-$29.43 & $-$21.36 & 1.00 \\ 
    $\mathrm{CeBiS_{2}O}$ & \underline{\textbf{94.89}} & N/A & N/A & 0.42 \\ 
    $\mathrm{Bi_{2}Sr_{2}CuO_{6}}$ & \underline{\textbf{127.33}} & 18.34 & 14.38 & 52.25 \\ 
    $\mathrm{TlSr_{2}CaCu_{2}O_{7}}$ & \underline{\textbf{74.64}} & 5.21 & 11.53 & 34.07 \\ 
    \hline
\end{tabular}
}
\end{table}

\begin{table}
\centering \small
\caption{Samples of proposed superconductors.}
\label{table__result_of_samples_of_exsiting_composition_based}
\resizebox{0.5\textwidth}{!}{
\begin{tabular}{c|c}
\hline
Method &  Samples \\
    \hline
    \multirow{4}{*}{KIAGO} & $\mathrm{Be_{2}Ca_{3}Cu_{3}Ce_{2}HgO_{9} (112.25 K)}$ \\  
     & $\mathrm{MgCa_{3}Cu_{3}Ba_{3}BiO_{9} (142.72 K)}$ \\  
     & $\mathrm{Ca_{4}Cu_{5}Sr_{4}NdN_{2}As_{5} (102.25 K)}$ \\  
     & $\mathrm{Ca_{5}Cu_{4}Sr_{4}CO_{10} (129.11 K)}$ \\ 
    \hline
    \multirow{4}{*}{SD} & $\mathrm{Ni_{0.99}LaO_{1.01}As_{0.99} (3.65 K)}$ \\  
     & $\mathrm{Ni_{0.82}Ge_{0.25}La_{0.97}C_{1.59}As_{0.44}Se_{0.14} (1.23 K)}$ \\  
     & $\mathrm{Ca_{0.28}Cu_{1.39}Sr_{2.04}Pb_{0.91}Bi_{1.15}O_{7.03} (76.03 K)}$ \\  
     & $\mathrm{Ca_{1.84}Sc_{0.17}Cu_{2.91}Ba_{2.05}HgO_{8.05} (127.46 K)}$ \\ 
    \hline
    \multirow{4}{*}{SD w/ CG} & $\mathrm{Be_{357.07}Fe_{0.31}Co_{0.21}Zn_{0.21}Ge_{0.93}La_{1.06} (8.41 K)}$ \\  
     & $\mathrm{Ca_{1.98}Cu_{3}Ba_{2.01}HgN_{0.2}O_{7.79} (125.64 K)}$ \\  
     & $\mathrm{Ca_{2.04}Cu_{3}Ba_{2.07}Hg_{1.09}O_{8.2} (126.48 K)}$ \\  
     & $\mathrm{CaCu_{2}Sr_{1.99}Hg_{0.77}Bi_{0.36}O_{7.03} (82.81 K)}$ \\ 
    \hline
    \multirow{4}{*}{C-ES} & $\mathrm{Ca_{2}Cu_{6}Sr_{4}O_{14} (103.89 K)}$ \\  
     & $\mathrm{CaCu_{3}Sr_{2}O_{7} (103.89 K)}$ \\  
     & $\mathrm{Ga_{4}Tl_{6}Bi_{18}O_{18}Si_{8}S_{36} (3.30 K)}$ \\  
     & $\mathrm{Ca_{6}Cu_{18}Sr_{12}O_{42} (103.89 K)}$ \\ 
    \hline
\end{tabular}
}
\end{table}

Interestingly, SuperDiff with CG does not necessarily generate higher-$T_c$ compounds than SuperDiff alone. Table~\ref{table__result_of_CG_vs_lr} shows how $T_c$ changes in guidance and denoising under different weights for guidance. During generation, high guidance weights raise $T_c$ in the guidance step but then revert it in the ILVR and denoising step. We attribute this to the $T_c$ distribution in SuperCon, where low- or moderate-$T_c$ compounds dominate (median: 12.5 K). As a result, extremely high $T_c$ values are treated as noise, prompting the model to restore them to more typical levels. Furthermore, larger weight for guidance cause a stronger mismatch with the training distribution, reducing the fraction of generated compositions that contain fewer than ten elements. This interplay of denoising and guidance likely hampers SuperDiff with CG's ability to reach stable, high-$T_c$ solutions. For the results without ILVR, please refer to Section~\ref{section__apx__superdiff}.

\begin{table}
\centering \small
\caption{Total changes in $T_c$ resulting from guidance, ILVR, and denoising during the 1000 steps in SuperDiff w/ CG. "Guide weight” denotes the weight for guidance. "Denoise $\Delta T_c$", "ILVR $\Delta T_c$" and “Guide $\Delta T_c$” represent the cumulative change in $T_c$ per step due to denoising, ILVR, or the guidance. “Sum” is the total of these values. "Screening ratio" denotes the ratio of the number of screened samples to the total number of samples.}
\label{table__result_of_CG_vs_lr}
\resizebox{0.5\textwidth}{!}{
\begin{tabular}{c|wc{1cm}wc{1cm}wc{1cm}wc{1cm}:wc{1cm}}
\hline
\begin{tabular}{c}Guide \\weight \\$w$\end{tabular}&  \begin{tabular}{c}Denoise\\ $\Delta T_c$(K) \end{tabular} &  \begin{tabular}{c}ILVR\\ $\Delta T_c$(K) \end{tabular} &  \begin{tabular}{c}Guide\\ $\Delta T_c$(K) \end{tabular} &  Sum (K) & \begin{tabular}{c} Screening\\ratio\end{tabular} \\
\hline
    - & 12.9 &    $-$3.4 &- &  9.5 &   0.08 \\
    1.0e-05 & 12.1 &    $-$2.5 &0.0 &  9.5 &   0.08 \\
    1.0e-04 & 12.6 &    $-$2.6 &0.0 & 10.0 &   0.08 \\
    1.0e-03 & 12.9 &    $-$3.7 &0.3 &  9.5 &   0.08 \\
    1.0e-02 & 11.9 &    $-$4.9 &2.6 &  9.7 &   0.09 \\
    1.0e-01 &  6.7 &   $-$22.0 &     25.3 &  9.9 &   0.08 \\
    1.0e+00 &$-$33.4 &  $-$254.1 &    309.0 & 21.5 &   0.00 \\
    1.0e+01 &$-$40.1 & $-$1658.2 &   1774.3 & 76.0 &   0.00 \\
    1.0e+02 &$-$18.2 & $-$1075.8 &   1141.9 & 48.0 &   0.00 \\
\hline
\end{tabular}
}
\end{table}

\subsection{Elemental substitution} \label{section__element_substitution}

In this section, we implemented elemental substitution to improve $T_c$. Here we replaced one metal element based on its oxidation state while retaining the rest of the composition. Specifically, we chose a single metal element from an existing superconductor and kept the remaining composition fixed.

We implement this approach in KIAGO by treating the preserved composition as a fixed vector $x_{const}$. We then randomly initialise the substituting element and apply a mask based on its oxidation state. For example, when substituting  $\mathrm{Y^{+3}}$, we only allow elements having a +3 oxidation state, such as gallium or aluminum. To achieve this, we used a mask that has one value on elements having a +3 oxidation state and set all others to zero. To simplify evaluation, we excluded the preserved elements from the mask. We use Adam optimiser with learning rate of 0.03 for KAIGO. Note that we did not perform conversion to integers on the substituting element, so we set $\beta=0$. By contrast, SuperDiff does not explicitly support elemental substitution, so we approximated it by conditioning the generation process with ILVR.

In addition to the screening described in Section~\ref{section___implementatin_details}, we assessed whether the intended elemental substitution was correctly carried out. First, we checked whether the preserved composition remains within 1\% of its original ratio. Second, we checked that the total composition of the newly substituted element (or elements) stays within 1\% of the original substituted metal's ratio. To simplify evaluation, we excluded the preserved elements from the substituted element candidates. We then evaluated the $T_c$ of compositions that pass both this substitution check and the previous screening.

Tables~\ref{table__result_of_element_substitution_prob}, \ref{table__result_of_element_substitution_neutrality}, and \ref{table__result_of_element_substitution_tc} summarize the probability of correct element evaluation, the charge-neutrality evaluation, and the resulting $T_c$ values, respectively. Notably, KIAGO and C-ES achieve 100\% correct substitutions (Table~\ref{table__result_of_element_substitution_prob}), indicating that these methods incorporate domain knowledge effectively. Consequently, as shown in Table~\ref{table__result_of_element_substitution_neutrality}, their proposed materials always satisfy charge neutrality. Moreover, KIAGO demonstrates high search efficiency, yielding the best results in most experiments (Table~\ref{table__result_of_element_substitution_tc}). SuperDiff, however, cannot reliably perform elemental substitution, indicating that generative models like SuperDiff are not well suited when strict domain knowledge must be enforced.

\begin{table}
\centering \small
\caption{Success rate of elemental substitution in proposed materials. We defined a successful elemental substitution as satisfying both of the following criteria: 1) the preserved composition remains within 1\% of its original ratio, and 2) the total composition of the newly substituted element (or elements) stays within 1\% of the original substituted metal's ratio.}
\label{table__result_of_element_substitution_prob}
\resizebox{0.5\textwidth}{!}{
\begin{tabular}{wc{1.9cm}|wc{1.0cm}:wc{0.8cm}wc{0.8cm}wc{0.8cm}wc{0.8cm}}
\hline
    \begin{tabular}{c}Base materials \\ from SuperCon\end{tabular} & \begin{tabular}{c}Substitute\\target\end{tabular} & KIAGO & SD & SD w/ CG &  C-ES \\
    \hline
    $\mathrm{CeFeAsF_{0.2}O_{0.8}}$ & $\mathrm{Ce^{+3}}$ & \underline{\textbf{1.00}} & 0.00 & 0.00 & \underline{\textbf{1.00}} \\ 
    $\mathrm{LaFeAsO_{1.0}}$ & $\mathrm{La^{+3}}$ & \underline{\textbf{1.00}} & 0.01 & 0.01 & \underline{\textbf{1.00}} \\ 
    $\mathrm{SrFe_{2}As_{2}}$ & $\mathrm{Sr^{+2}}$ & \underline{\textbf{1.00}} & 0.01 & 0.01 & \underline{\textbf{1.00}} \\ 
    $\mathrm{Bi_{2}CaSr_{2}Cu_{2}O_{8}}$ & $\mathrm{Bi^{+3}}$ & \underline{\textbf{1.00}} & 0.23 & 0.22 & \underline{\textbf{1.00}} \\ 
    $\mathrm{CeNiC_{2}}$ & $\mathrm{Ce^{+4}}$ & \underline{\textbf{1.00}} & 0.00 & 0.00 & \underline{\textbf{1.00}} \\ 
    $\mathrm{LaNiC_{2}}$ & $\mathrm{La^{+3}}$ & \underline{\textbf{1.00}} & 0.01 & 0.01 & \underline{\textbf{1.00}} \\ 
    $\mathrm{MgCoNi_{3}}$ & $\mathrm{Co^{+2}}$ & \underline{\textbf{1.00}} & 0.00 & 0.00 & \underline{\textbf{1.00}} \\ 
    $\mathrm{RuSr_{2}GdCu_{2}O_{8}}$ & $\mathrm{Sr^{+2}}$ & \underline{\textbf{1.00}} & 0.03 & 0.03 & \underline{\textbf{1.00}} \\ 
    $\mathrm{RuSr_{2}YCu_{2}O_{8}}$ & $\mathrm{Y^{+3}}$ & \underline{\textbf{1.00}} & 0.03 & 0.03 & \underline{\textbf{1.00}} \\ 
    $\mathrm{Y_{2}Fe_{3}Si_{5}}$ & $\mathrm{Y^{+3}}$ & \underline{\textbf{1.00}} & 0.00 & 0.00 & \underline{\textbf{1.00}} \\ 
    $\mathrm{YIrSi}$ & $\mathrm{Y^{+3}}$ & \underline{\textbf{1.00}} & 0.00 & 0.00 & \underline{\textbf{1.00}} \\ 
    \hline
\end{tabular}
}
\end{table}

\begin{table}[t]
\centering \small
\caption{Success rate with respect to charge neutrality in proposed materials resulting from elemental substitution.}
\label{table__result_of_element_substitution_neutrality}
\resizebox{0.5\textwidth}{!}{
\begin{tabular}{wc{1.9cm}|wc{1.0cm}:wc{0.8cm}wc{0.8cm}wc{0.8cm}wc{0.8cm}}
\hline
    \begin{tabular}{c}Base materials \\ from SuperCon\end{tabular} & \begin{tabular}{c}Substitute\\target\end{tabular} & KIAGO & SD & SD w/ CG &  C-ES \\
    \hline
    $\mathrm{CeFeAsF_{0.2}O_{0.8}}$ & $\mathrm{Ce^{+3}}$ & \underline{\textbf{1.00}} & 0.03 & 0.02 & \underline{\textbf{1.00}} \\ 
    $\mathrm{LaFeAsO_{1.0}}$ & $\mathrm{La^{+3}}$ & \underline{\textbf{1.00}} & 0.02 & 0.02 & \underline{\textbf{1.00}} \\ 
    $\mathrm{SrFe_{2}As_{2}}$ & $\mathrm{Sr^{+2}}$ & \underline{\textbf{1.00}} & 0.03 & 0.04 & \underline{\textbf{1.00}} \\ 
    $\mathrm{Bi_{2}CaSr_{2}Cu_{2}O_{8}}$ & $\mathrm{Bi^{+3}}$ & \underline{\textbf{1.00}} & 0.01 & 0.01 & \underline{\textbf{1.00}} \\ 
    $\mathrm{CeNiC_{2}}$ & $\mathrm{Ce^{+4}}$ & \underline{\textbf{1.00}} & 0.07 & 0.06 & \underline{\textbf{1.00}} \\ 
    $\mathrm{LaNiC_{2}}$ & $\mathrm{La^{+3}}$ & \underline{\textbf{1.00}} & 0.03 & 0.04 & \underline{\textbf{1.00}} \\ 
    $\mathrm{MgCoNi_{3}}$ & $\mathrm{Co^{+2}}$ & \underline{\textbf{1.00}} & 0.71 & 0.66 & \underline{\textbf{1.00}} \\ 
    $\mathrm{RuSr_{2}GdCu_{2}O_{8}}$ & $\mathrm{Sr^{+2}}$ & \underline{\textbf{1.00}} & 0.05 & 0.06 & \underline{\textbf{1.00}} \\ 
    $\mathrm{RuSr_{2}YCu_{2}O_{8}}$ & $\mathrm{Y^{+3}}$ & \underline{\textbf{1.00}} & 0.08 & 0.07 & \underline{\textbf{1.00}} \\ 
    $\mathrm{Y_{2}Fe_{3}Si_{5}}$ & $\mathrm{Y^{+3}}$ & \underline{\textbf{1.00}} & 0.18 & 0.43 & \underline{\textbf{1.00}} \\ 
    $\mathrm{YIrSi}$ & $\mathrm{Y^{+3}}$ & \underline{\textbf{1.00}} & 0.28 & 0.37 & \underline{\textbf{1.00}} \\ 
    \hline
\end{tabular}
}
\end{table}

\begin{table}
\centering \small
\caption{Differences in $T_c$ (K) between the average $T_c$ of the highest Top-30 proposed superconductors and the base superconductors in experiments of elemental substitution. After screening 4096 samples, the $T_c$ prediction model was used to calculate $T_c$. 'N/A' indicates that none of samples passed the screening.}
\label{table__result_of_element_substitution_tc}
\resizebox{0.5\textwidth}{!}{
\begin{tabular}{wc{1.9cm}|wc{1.0cm}:wc{0.8cm}wc{0.8cm}wc{0.8cm}wc{0.8cm}}
\hline
    \begin{tabular}{c}Base materials \\ from SuperCon\end{tabular} & \begin{tabular}{c}Substitute\\target\end{tabular} & \begin{tabular}{c}KIAGO\\Top-30\\ $\Delta T_c$(K) \end{tabular} & \begin{tabular}{c}SD\\Top-30\\ $\Delta T_c$(K) \end{tabular} & \begin{tabular}{c}SD w/ CG\\Top-30\\ $\Delta T_c$(K) \end{tabular} & \begin{tabular}{c}C-ES\\Top-30\\ $\Delta T_c$(K) \end{tabular} \\
    \hline
    $\mathrm{CeFeAsF_{0.2}O_{0.8}}$ & $\mathrm{Ce^{+3}}$ & \underline{\textbf{14.53}} & N/A & N/A & 9.99 \\ 
    $\mathrm{LaFeAsO_{1.0}}$ & $\mathrm{La^{+3}}$ & \underline{\textbf{35.62}} & 1.24 & N/A & 29.45 \\ 
    $\mathrm{SrFe_{2}As_{2}}$ & $\mathrm{Sr^{+2}}$ & \underline{\textbf{29.35}} & N/A & N/A & 13.78 \\ 
    $\mathrm{Bi_{2}CaSr_{2}Cu_{2}O_{8}}$ & $\mathrm{Bi^{+3}}$ & \underline{\textbf{14.61}} & 0.15 & $-$0.27 & 6.42 \\ 
    $\mathrm{CeNiC_{2}}$ & $\mathrm{Ce^{+4}}$ & \underline{\textbf{17.30}} & N/A & N/A & 6.26 \\ 
    $\mathrm{LaNiC_{2}}$ & $\mathrm{La^{+3}}$ & \underline{\textbf{25.24}} & N/A & 0.11 & 7.55 \\ 
    $\mathrm{MgCoNi_{3}}$ & $\mathrm{Co^{+2}}$ & N/A & N/A & N/A & \underline{\textbf{-2.41}} \\ 
    $\mathrm{RuSr_{2}GdCu_{2}O_{8}}$ & $\mathrm{Sr^{+2}}$ & $-$33.66 & $-$0.23 & 0.93 & \underline{\textbf{3.70}} \\ 
    $\mathrm{RuSr_{2}YCu_{2}O_{8}}$ & $\mathrm{Y^{+3}}$ & \underline{\textbf{46.16}} & $-$0.89 & $-$1.58 & 37.60 \\ 
    $\mathrm{Y_{2}Fe_{3}Si_{5}}$ & $\mathrm{Y^{+3}}$ & \underline{\textbf{7.56}} & N/A & N/A & 1.45 \\ 
    $\mathrm{YIrSi}$ & $\mathrm{Y^{+3}}$ & 5.08 & 3.42 & N/A & \underline{\textbf{5.31}} \\ 
    \hline
\end{tabular}
}
\end{table}

Next, we compare the highest-$T_c$ compounds in the SuperCon dataset that have undergone the same elemental substitution with the compounds proposed by KIAGO. Table~\ref{table__result_of_element_substitution_tc_vs_database} shows that, in most element-substitution experiments, KIAGO proposes materials with higher $T_c$ than any element-substituted materials in the SuperCon dataset. This result highlights the potential of our approach to surpass known substitution strategies and discover more promising superconductors.

\begin{table*}
\centering \small
\caption{Comparison between materials proposed by KIAGO and the highest-$T_c$ compounds from the SuperCon dataset under the same element-substitution conditions. All $T_c$ values are prediction values by our $T_c$ prediction model. Blue elements indicate the substituted elements.}
\label{table__result_of_element_substitution_tc_vs_database}
\resizebox{\textwidth}{!}{
\begin{tabular}{l:wc{1.0cm}|cc}
\hline
    \begin{tabular}{c}Base materials \\ from SuperCon\end{tabular} & \begin{tabular}{c}Substitute\\target\end{tabular} &  Proposed Samples & Best in SuperCon \\
    \hline
    $\mathrm{CeFeAsF_{0.2}O_{0.8}}$ \  (39.8 K) & $\mathrm{Ce^{+3}}$ & $\mathrm{}\textcolor{blue}{\mathbf{Ti_{0.1807}Sm_{0.8193}}}\mathrm{FeAsF_{0.2}O_{0.8}}$ (\underline{\textbf{57.08}} K) & $\mathrm{}\textcolor{blue}{\mathbf{Gd}}\mathrm{FeAsF_{0.2}O_{0.8}}$ \  (48.3 K) \\ 
    $\mathrm{LaFeAsO_{1.0}}$ \  (13.4 K) & $\mathrm{La^{+3}}$ & $\mathrm{}\textcolor{blue}{\mathbf{Nd_{0.6077}Gd_{0.3239}Tb_{0.0684}}}\mathrm{FeAsO}$ (\underline{\textbf{53.64}} K) & $\mathrm{}\textcolor{blue}{\mathbf{Sm_{0.65}Th_{0.35}}}\mathrm{FeAsO}$ \  (52.9 K) \\ 
    $\mathrm{SrFe_{2}As_{2}}$ \  (17.3 K) & $\mathrm{Sr^{+2}}$ & $\mathrm{}\textcolor{blue}{\mathbf{Mg_{0.2089}V_{0.0183}La_{0.243}Nd_{0.251}Eu_{0.2014}Dy_{0.0774}}}\mathrm{Fe_{2}As_{2}}$ (\underline{\textbf{47.59}} K) & $\mathrm{}\textcolor{blue}{\mathbf{Ba_{0.66}K_{0.34}}}\mathrm{Fe_{2}As_{2}}$ \  (38.6 K) \\ 
    $\mathrm{Bi_{2}CaSr_{2}Cu_{2}O_{8}}$ \  (81.0 K) & $\mathrm{Bi^{+3}}$ & $\mathrm{}\textcolor{blue}{\mathbf{Hf_{0.2381}Pt_{0.5835}Bi_{1.1784}}}\mathrm{CaSr_{2}Cu_{2}O_{8}}$ (\underline{\textbf{95.88}} K) & $\mathrm{}\textcolor{blue}{\mathbf{Bi_{1.2}Pb_{0.8}}}\mathrm{CaSr_{2}Cu_{2}O_{8}}$ \  (92.9 K) \\ 
    $\mathrm{CeNiC_{2}}$ \  (3.0 K) & $\mathrm{Ce^{+4}}$ & $\mathrm{}\textcolor{blue}{\mathbf{N_{0.1514}S_{0.1395}V_{0.086}Zr_{0.2171}Rh_{0.3626}Pr_{0.0434}}}\mathrm{NiC_{2}}$ (\underline{\textbf{20.91}} K) & $\mathrm{}\textcolor{blue}{\mathbf{Th_{0.1}Y_{0.9}}}\mathrm{NiC_{2}}$ \  (6.6 K) \\ 
    $\mathrm{LaNiC_{2}}$ \  (2.4 K) & $\mathrm{La^{+3}}$ & $\mathrm{}\textcolor{blue}{\mathbf{Zr_{0.1793}Rh_{0.4886}Sb_{0.293}Pr_{0.0391}}}\mathrm{NiC_{2}}$ (\underline{\textbf{27.82}} K) & $\mathrm{}\textcolor{blue}{\mathbf{Th_{0.1}Y_{0.9}}}\mathrm{NiC_{2}}$ \  (6.6 K) \\ 
    $\mathrm{MgCoNi_{3}}$ \  (7.4 K) & $\mathrm{Co^{+2}}$ & N/A & $\mathrm{Mg}\textcolor{blue}{\mathbf{Cu}}\mathrm{Ni_{3}}$ \  (\underline{\textbf{7.6 K}}) \\ 
    $\mathrm{RuSr_{2}GdCu_{2}O_{8}}$ \  (33.7 K) & $\mathrm{Sr^{+2}}$ & $\mathrm{Ru}\textcolor{blue}{\mathbf{N_{0.4994}Br_{0.5585}Nd_{0.5374}Er_{0.4047}}}\mathrm{GdCu_{2}O_{8}}$ (0.00 K) & $\mathrm{Ru}\textcolor{blue}{\mathbf{Ba_{0.4}Sr_{1.6}}}\mathrm{GdCu_{2}O_{8}}$ \  (40.7 K) \\ 
    $\mathrm{RuSr_{2}YCu_{2}O_{8}}$ \  (34.4 K) & $\mathrm{Y^{+3}}$ & $\mathrm{RuSr_{2}}\textcolor{blue}{\mathbf{Ti_{0.3309}Nd_{0.1602}W_{0.2576}Bi_{0.2513}}}\mathrm{Cu_{2}O_{8}}$ (\underline{\textbf{80.61}} K) & $\mathrm{RuSr_{2}}\textcolor{blue}{\mathbf{Ca_{0.1}Gd_{0.9}}}\mathrm{Cu_{2}O_{8}}$ \  (60.4 K) \\ 
    $\mathrm{Y_{2}Fe_{3}Si_{5}}$ \  (2.3 K) & $\mathrm{Y^{+3}}$ & $\mathrm{}\textcolor{blue}{\mathbf{Pr_{0.4884}Sm_{0.1106}Gd_{0.3835}Dy_{0.4279}Ho_{0.3464}Hf_{0.2432}}}\mathrm{Fe_{3}Si_{5}}$ (\underline{\textbf{9.90}} K) & $\mathrm{}\textcolor{blue}{\mathbf{Lu_{2}}}\mathrm{Fe_{3}Si_{5}}$ \  (5.3 K) \\ 
    $\mathrm{YIrSi}$ \  (2.8 K) & $\mathrm{Y^{+3}}$ & $\mathrm{}\textcolor{blue}{\mathbf{Co_{0.1167}La_{0.0731}Ce_{0.1664}Sm_{0.2572}Eu_{0.2764}Tb_{0.1102}}}\mathrm{IrSi}$ (\underline{\textbf{10.15}} K) & $\mathrm{}\textcolor{blue}{\mathbf{Th}}\mathrm{IrSi}$ \  (6.9 K) \\ 
    \hline
\end{tabular}
}
\end{table*}

In this section, we limit our discussion to single-element substitution. However, our method can also support multi-element substitution. For example, in $\mathrm{Ti_2O_4}$, two $\mathrm{Ti}^{4+}$ atoms and one $\mathrm{O}^{2-}$ atom contribute a total charge of $+6$. This can be replaced by two $\mathrm{X}^{3+}$ atoms, resulting in a composition like $\mathrm{X_2O_3}$. Here, $\mathrm{X}$ denotes any element with a $+3$ oxidation state. Such substitutions are feasible as long as the total charge is preserved, and our oxidation-state-based masking mechanism can accommodate them.

\subsection{Proposing novel hydride superconductors}

In this section, we focused on proposing novel hydride superconductors (HSC) . Many known HSC are binary or ternary systems containing hydrogen and just one or two other elements, with hydrogen comprising a large fraction of the composition. Thus, we constrained KIAGO to compositions that have at least 40\% hydrogen to expand the space around existing materials, form binary or ternary compounds, and possess a total atom count of 15 or fewer.
HSC are often tested under high pressure, where Pauling's electronegativity rules may not hold. For instance, $\mathrm{LaH_{10}}$ has been experimentally confirmed but fails SMACT-based screening for electronegativity and charge neutrality, because SMACT assumes each element has a single fixed valence. Then, we assumed each atom of the same element could adopt different valences. For example, hydrogen can be both $+1$ and $-1$, making $\mathrm{LaH_{10}}= \{ \mathrm{La^{+2}, H^{-1}\times6, H^{+1}\times4} \}$, which is thus electrically neutral. However, allowing multiple valences can lead to a combinatorial explosion, so we imposed a maximum total of 15 atoms per composition. We used these criteria (charge neutrality under variable valences, ternary or binary composition, and a total of 15 or fewer atoms) as a replacement for SMACT-based screening.

For initialisation for KIAGO, we began with HSC from the SuperCon dataset. With a probability of 0.29, we replaced elements of its composition with different elements chosen according to their occurrence frequencies in SuperCon. We then selected random elements with random compositional ratios (from 0.0 to 0.03) for those elements, normalised the resulting composition, and used it as the initialisation. We also set $\{ N_\mathrm{unit}\}=\{1,2,...,15\}$.
 
In Table~\ref{table__result_hydride_supercon}, we present the average $T_c$ of the top five proposed hydride superconductors. Compared with other methods, KIAGO efficiently generates hydride superconductors. Table\ref{table__result_of_valid_hsc_rate} shows the probability of proposing materials that meet specific criteria—namely, having at least 40\% hydrogen content, three or fewer elements, and a total atom count of 15 or below. These results indicate that KIAGO not only explores the search space efficiently but also strictly adheres to the specified constraints.

Table~\ref{table__superconducting_materials} lists HSC proposed by KIAGO. Notably, KIAGO also proposed materials made of the same elements as those in known compounds from the SuperCon dataset. In addition, it suggested materials that are not in SuperCon dataset but have been reported in other literature.

\begin{table}
\centering \small
\caption{ Average $T_c$ for the Top-5 proposed superconductors}
\label{table__result_hydride_supercon}
\resizebox{0.5\textwidth}{!}{
\begin{tabular}{c|wc{1cm}wc{1cm}wc{1cm}wc{1cm}}
\hline
    \begin{tabular}{c}Base materials \\ from SuperCon\end{tabular} & \begin{tabular}{c}KIAGO\\Top-5\\$T_c$(K) \end{tabular} & \begin{tabular}{c}SD\\Top-5\\$T_c$(K) \end{tabular} & \begin{tabular}{c}SD w/ CG\\Top-5\\$T_c$(K) \end{tabular} & \begin{tabular}{c}C-ES\\Top-5\\$T_c$(K) \end{tabular} \\
    \hline
    $\mathrm{PdH}$ & \underline{\textbf{3.58}} & 0.00 & 0.10 & 3.03 \\ 
    $\mathrm{PtH}$ & \underline{\textbf{3.13}} & 0.00 & 0.00 & 3.07 \\ 
    $\mathrm{LaH_{10}}$ & \underline{\textbf{3.72}} & 0.00 & 0.00 & 0.00 \\ 
    $\mathrm{H_{2}S}$ & \underline{\textbf{3.10}} & 0.14 & 0.14 & 0.00 \\ 
    $\mathrm{H_{4}Si}$ & \underline{\textbf{3.13}} & 0.00 & 0.00 & 0.00 \\ 
    \hline
\end{tabular}
}
\end{table}

\begin{table}
\centering \small
\caption{Ratio of proposed materials satisfying the following three conditions: (1) Hydrogen($\mathrm{H}$) content is 40\% or more, (2) composed of three or fewer elements, and (3) 15 atoms or less.}
\label{table__result_of_valid_hsc_rate}
\begin{tabular}{l|cccc}
\hline
    \begin{tabular}{c}Base materials \\ from SuperCon\end{tabular} & KIAGO & SD & SD w/ CG &  C-ES \\
    \hline
    $\mathrm{PdH}$ & \underline{\textbf{1.00}} & 0.03 & 0.02 & \underline{\textbf{1.00}} \\ 
    $\mathrm{PtH}$ & \underline{\textbf{1.00}} & 0.03 & 0.03 & \underline{\textbf{1.00}} \\ 
    $\mathrm{LaH_{10}}$ & \underline{\textbf{1.00}} & 0.02 & 0.02 & \underline{\textbf{1.00}} \\ 
    $\mathrm{H_{2}S}$ & \underline{\textbf{1.00}} & 0.04 & 0.03 & \underline{\textbf{1.00}} \\ 
    $\mathrm{H_{4}Si}$ & \underline{\textbf{1.00}} & 0.03 & 0.02 & \underline{\textbf{1.00}} \\ 
    \hline
\end{tabular}
\end{table}

\begin{table}
\centering \small
\caption{Candidates of hydride superconductor proposed by KIAGO. The The 'Similar formula in Refs' and '$T_c$ in Refs' columns show the composition and $T_c$ of superconductors experimentally confirmed or calculated by DFT in other studies, respectively, which are composed of the same elements as the proposed materials.}
\label{table__superconducting_materials}
\resizebox{0.5\textwidth}{!}{
\begin{tabular}{wc{1.0cm}wc{1.0cm}:wc{1.0cm}wc{1.0cm}:wc{1.0cm}wc{1.0cm}}
\hline
  \begin{tabular}{c}Proposed\\formula\end{tabular} &  \begin{tabular}{c}Predicted\\$T_c$ (K)\end{tabular} & \begin{tabular}{c}Similar\\formula\\in dataset\end{tabular} &  \begin{tabular}{c}$T_c$ (K) in\\ SuperCon\end{tabular} & \begin{tabular}{c}Similar\\formula\\in Refs\end{tabular} & \begin{tabular}{c}$T_c$ (K)\\in Refs\end{tabular} \\
\hline
    $\mathrm{SiH}$ &   0.7 &  $\mathrm{SiH_4}$ &   17.0 &  - &  - \\
\hdashline 
    $\mathrm{ZrH}$ &   0.5 &   - & - &   $\mathrm{ZrH3}$ & $6.7^\mathrm{EXP}$\cite{xie2020superconducting} \\
$\mathrm{H_9Mo_5}$ &   6.7 &   - & - &  $\mathrm{MoH_6}$ & $80^\mathrm{DFT}$\cite{10.1063/5.0005873} \\
$\mathrm{Be_2H_5}$ &   2.3 &   - & - &  $\mathrm{BeH_3}$ &   $181^\mathrm{DFT}$\cite{gao2024prediction} \\
$\mathrm{Bi_7H_5}$ &   3.8 &   - & - &  $\mathrm{BiH_4}$ & $91^\mathrm{DFT}$\cite{shan2024molecular} \\
$\mathrm{Ge_7H_5}$ &   2.3 &   - & - &  $\mathrm{GeH_4}$ &  $64^\mathrm{DFT}$\cite{gao2008superconducting} \\
\hline
\end{tabular}
}
\end{table}

\section{Limitation}

Our method relies heavily on the accuracy of the prediction models. Although our current $T_c$ predictor achieves competitive performance compared with other methods (see Section~\ref{sectin__tc_scores}), the predicted $T_c$ values for the proposed materials inevitably contain some error. In addition, since the SuperCon dataset lacks pressure information—crucial for hydride superconductors—our model cannot address pressure effects, which could pose another limitation. Improving the prediction model's accuracy must involve ensemble methods~\cite{taheri2022prediction}, better model architectures, or enhanced datasets. Importantly, our method does not depend on any specific model architecture. Given the rapid pace of machine-learning advances, more accurate models will likely become available soon, and substituting them into our framework should alleviate current limitations. Additionally, datasets are also improving at a fast rate, offering further opportunities for refinement.

\section*{Conclusion}
In this paper, we introduced KIAGO, a gradient-based method for proposing high-$T_c$ superconductors that unify domain knowledge with efficient computational strategies. Unlike Classifier Guidance-based generative models, KIAGO does not require to train additional generative models, making it a more straightforward solution. By initialising the optimisation from promising superconductors, we mitigate the risk of converging to poor local minima—an issue often encountered in gradient-based methods—and achieve higher optimisation efficiency. A key strength of KIAGO lies in its ability to incorporate diverse domain knowledge via masking. We demonstrated this by precisely controlling elemental substitutions and restricting our search to hydride superconductors. These results underscore the adaptability of KIAGO: it not only capitalizes on existing knowledge, like traditional doping strategies but also explores a broader chemical space more effectively than previous approaches. Overall, KIAGO paves the way for discovering new materials by exploiting domain knowledge and machine learning's scalability. This synergy has the potential to accelerate advancements in high-$T_c$ superconductivity and beyond, offering a robust framework for rapid and adaptive materials design.

\section*{Conflicts of interest}
There are no conflicts to declare.

\section*{Acknowledgement}
The authors gratefully acknowledge support from the Doctoral Student Special Incentives Program, Graduate School of Engineering, The University of Tokyo (SEUT-RA).

\section*{Data availability}
The code used to implement the optimization framework in this study is available at Zenodo: \url{https://doi.org/10.5281/zenodo.15477853}, which archives the GitHub repository at \url{https://github.com/AkiraTOSEI/KIAGO}.  This version corresponds to the release v0.1.0, accessed on 20 May 2025. Trained models for the $T_c$ predictor and formation energy predictor are also provided in this repository. The SuperCon dataset is available at \url{https://mdr.nims.go.jp/collections/5712mb227}. The ElemNet training data can be obtained from \url{http://cucis.ece.northwestern.edu/projects/DataSets/ElemNet/data.tar.gz}, and the COD database can be accessed at \url{https://www.crystallography.net/cod/}.

%%%END OF MAIN TEXT%%%


\balance


%%%REFERENCES%%%
\bibliography{reference} %You need to replace "rsc" on this line with the name of your .bib file
\bibliographystyle{unsrt} %the RSC's .bst file
%
%
%
%
% APPENDIX
%
%
%
%
\clearpage
\newpage

\appendix

% ← ここで図表の番号フォーマットを設定する（順番が重要！）
\renewcommand{\thesection}{S.\arabic{section}}                      % セクション番号 S.1, S.2,...
\renewcommand{\thefigure}{\Alph{section}.\arabic{figure}}           % 図番号 A.1, A.2,...
\renewcommand{\thetable}{\Alph{section}.\arabic{table}}             % 表番号 A.1, A.2,...
\numberwithin{equation}{section}                                   % 式番号 S.1, S.2,...
\numberwithin{figure}{section}
\numberwithin{table}{section}

\section{$T_c$ prediction model} \label{section__tc_prediction_model}

\begin{figure*}[t]
\centering
\includegraphics[width=\linewidth]{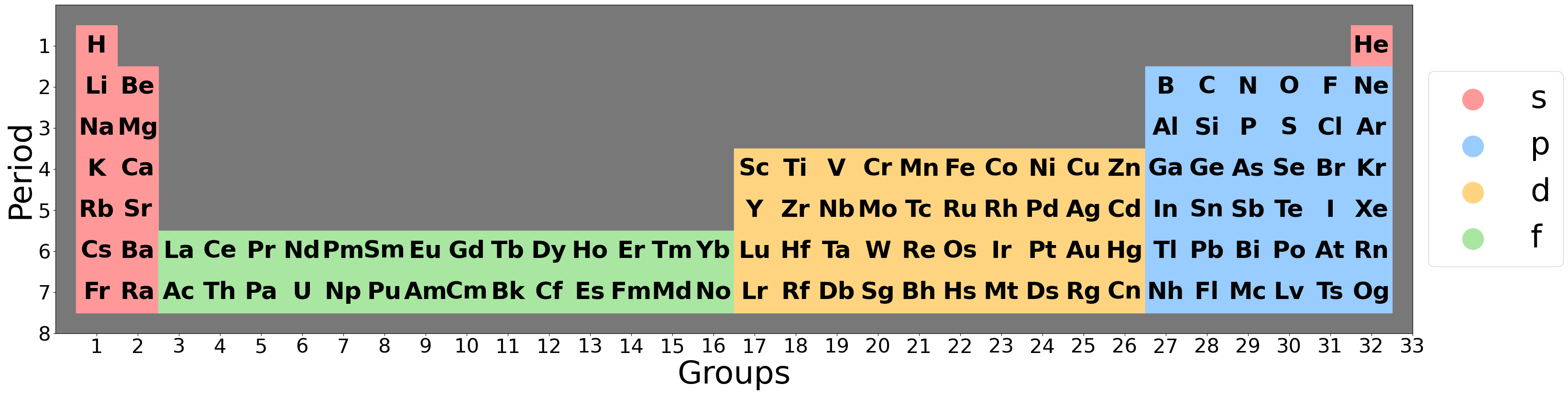}
\caption{Periodic table, coloured according to the type of orbital occupied by the outermost electron.}
\label{fgr:periodic_table}
\end{figure*}

\begin{figure}[h]
\centering
\includegraphics[width=\linewidth]{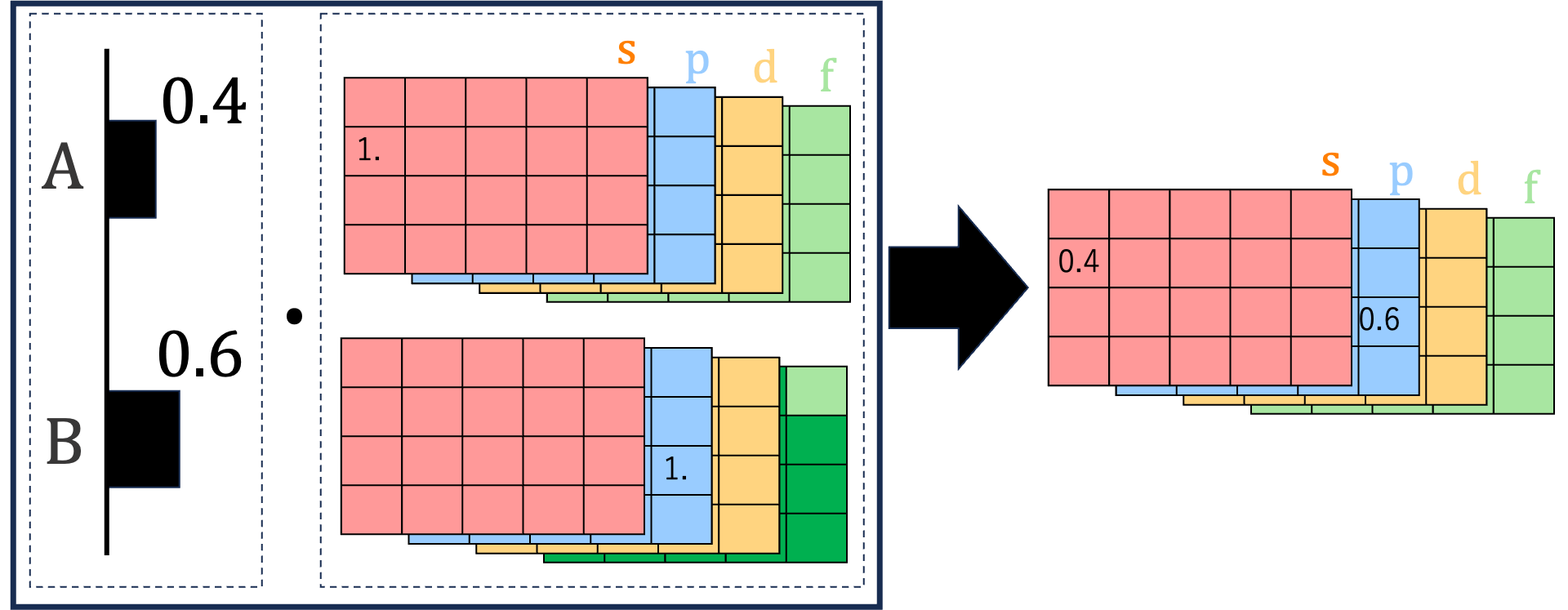}
\caption{Example: Representation of the composition $\mathrm{A_2B_3}$}
\label{fgr:representation_composition}
\end{figure}

We trained a deep learning model $f_{T_c}$ to predict the critical temperature ($T_c$) from composition, utilizing the SuperCon dataset as the source of superconductor data. Konno \textit{et al.}~\cite{konno2021deep} reported that training only on the SuperCon dataset increases false positives (i.e., mistakenly predicting non-superconductors as superconductors), so we added the Crystallography Open Database (COD)~\cite{vaitkus2023workflow,merkys2023graph,vaitkus2021validation,quiros2018using,merkys2016cod,gravzulis2015computing,gravzulis2012crystallography,gravzulis2009crystallography,downs2003american} as a dataset of non-superconductors to training data. We trained the ResNet18 regression model using Adam optimiser ~\cite{kingma2014adam} with a learning rate of 0.0001 and a batch size of 1024. Given that $T_c$ $\geq 0$, we employed the Rectified Linear Unit (ReLU) as the activation function at the output layer.

\subsection{Dataset}

We used the SuperCon dataset for superconductor materials. For preprocessing, we excluded substances whose compositions were not quantitatively specified, such as those denoted by variables (e.g., x, y, z). We also used the Crystallography Open Database (COD) for the non-superconductor materials. Furthermore, compositions identical to those in the SuperCon data were excluded from the COD data.

For both SuperCon and COD, compositions were represented by 118-dimensional (from $\mathrm{H}$ to $\mathrm{Og}$) distribution vectors $x$, normalised to sum to 1. The critical temperature ($T_c$) was used as the target variable $y$, setting $y$=0 for non-superconducting materials from the COD dataset and using the actual $T_c$ values from the SuperCon dataset. For substances with identical compositional ratios but different $T_c$ values, we calculated the mean $T_c$ value. Consequently, we obtained 14,410 data points for the SuperCon dataset and 373,901 data points for the COD dataset. Materials with the same combination of elements may cause data leakage if they are split across different data partitions. Therefore, we ensured that materials with the same combination of elements were placed in the same data partition during the data split. We then randomly split these into training, validation, and test sets in the ratios of 0.80:0.05:0.15, respectively.

However, we were concerned about the model's accuracy due to an imbalance in the training data resulting from a smaller number of superconductors compared to non-superconductors. Therefore, to address this imbalance, we augmented the training data by increasing the number of instances from the SuperCon dataset by a factor of 25 through replication.

\subsection{Element representation}

As element features $\mathbf{a}$, we utilized flag representations for the s, p, d, and f electron orbitals as embedded in the periodic table~\cite{konno2021deep}. Each element $i$ has a feature tensor $\mathbf{a}$ with dimensions of $4\times7\times32$, representing the channels for the s, p, d, and f electron orbitals, periods, and groups. Note that, due to the extension of lanthanides and actinides in the group direction, the third dimension (group) extends to 32, not 18. For any element $i$, its feature $\mathbf{a}^i$ assigns a value of 1 at the position in the periodic table corresponding to the outermost electron's channel and 0 values for all other positions.

\begin{gather}
\mathbf{A}_i = \mathbf{a}^i,\\ \mathbf{A} \in \{0, 1\}^{118 \times 4\times 7\times 32},\\
\quad i \in \{\mathbf{H},\mathbf{He},...,\mathbf{Og} \}\\
\end{gather}

\begin{gather}
\mathbf{A}_{ijkl} = \begin{cases} 1 & \left.  \begin{aligned}
    \text{if atom } i  \text{ has its outermost electrons}\\ \text{, family (group),and period }\\
    \text{corresponding to } j, k, \text{ and } l.
\end{aligned}  \right.   \\0 & \text{otherwise}\end{cases}
\end{gather}

For instance, the hydrogen atom, which has a 1s electron in its outermost shell and belongs to the first period and group, is represented by the feature $\mathbf{a}^\mathrm{H}$. This feature has a value of one at position $(1,1,1)$ and zeroes elsewhere. Similarly, the feature for chlorine, $\mathbf{a}^\mathrm{Cl}$, which contains 3p electrons in its outer shell and is located in the third period and seventeenth group, has a value of 1 at position $(2,3,31)$ and zeros elsewhere. Fig.~\ref{fgr:periodic_table} illustrates the corresponding channel of the outermost electron for each element. The composition's representation is calculated by computing the product of the composition vector and the element features $\mathbf{a}$, as shown in Fig.~\ref{fgr:representation_composition}.

\subsection{Prediction scores} \label{sectin__tc_scores}

Since previous methods were evaluated only on SuperCon data, we first evaluated our model solely on the SuperCon dataset (Table~\ref{tbl:tc_prediction_comparision}). Note that it is not a direct comparison due to differences in dataset dividing. The previous studies did not describe whether they consider the combination of elements in the data split. We assume that the previous studies did not perform such data splitting. Our model was the only one trained on both SuperCon and COD datasets. Our model demonstrated competitive results compared to other methods when using only SuperCon data. The model trained using COD and considering the combination of elements in the data split scored worse than the model trained only on SuperCon. However, as we will discuss in Section~\ref{sectin__reducing_false_postive}, it was found that using COD is important to reduce false positives. Therefore, we adopted this model.

\begin{table}[H]
\centering
\caption{Results on SuperCon data and comparison with other methods.}
\label{tbl:tc_prediction_comparision}
\resizebox{0.5\textwidth}{!}{
\begin{tabular}{c|cccc}
\hline
    Method &  Train Dataset & Elem Div & MAE &  R2 \\
\hline
   Stanev \textit{et al.}~\cite{stanev2018machine} &  SuperCon & False? &- &  0.88 \\
Zeng \textit{et al.}~\cite{zeng2019atom} &  SuperCon & False? &  4.21 &  0.97 \\
   Dan \textit{et al.}~\cite{dan2020computational} &  SuperCon & False? &- & 0.907 \\
Taheri \textit{et al.}~\cite{taheri2022prediction} &  SuperCon & False? &- & 0.927 \\ \hdashline
 Ours &  SuperCon &  False &  4.90 &  0.92 \\
 Ours & SuperCon + COD &True &  7.84 &  0.80 \\
\hline
\end{tabular}
}
\end{table}

\FloatBarrier

\subsection{Reducing false positives} \label{sectin__reducing_false_postive}

Here, we addressed the issue of false positives, that is, non-superconductors incorrectly predicted to be superconductors. Konno \textit{et al.}~\cite{konno2021deep} stated that models trained exclusively on the SuperCon dataset tend to produce a significant number of false positives, although this claim has not been numerically demonstrated. Hence, we numerically assessed how incorporating the COD dataset, which consists of non-superconductors, impacts the rate of false positives.

We performed this assessment using our model, a ResNet18 regression model and adopted the precision score as our metric. The precision score is defined as \(\mathrm{tp}/(\mathrm{tp}+\mathrm{fp})\), where tp (true positive) denotes the probability of accurately predicting a superconductor (y>0) as superconductor (i.e., \(f_{T_c}(x)>0\)), and fp (false positive) represents the probability of mistakenly predicting a non-superconductor (y=0) as a superconductor (i.e., \(f_{T_{c}}(x)>0\)). 

Table~\ref{tbl:false_positive} provides a comparison of precision scores between models trained on the SuperCon dataset alone and those trained on both SuperCon and COD datasets. It is evident that models trained with the inclusion of the COD dataset exhibit a significantly higher precision score compared to those trained exclusively on the SuperCon dataset. This outcome underscores the necessity of integrating non-superconductors into the training process for superconductor material design tasks, where non-superconductors might emerge as potential candidates.

\begin{table}
\centering
\caption{Precision scores with respect to the training set.}
\label{tbl:false_positive}
\begin{tabular}{cc}
\hline % To generate a thicker line than \hline
 Training data & Precision score \\
\hline
SuperCon& 0.065\\
SuperCon+COD& \textbf{0.804}\\
\hline
\end{tabular}
\end{table}

\section{Formation energy prediction model}  \label{section__elemnet}
In this study, we employed ElemNet~\cite{jha2018elemnet} as the formation energy prediction model. Although its original implementation (\url{https://github.com/NU-CUCIS/ElemNet}) use TensorFlow 1.x, we require gradient information in PyTorch. Thus, we re-implemented the model in PyTorch. We follow the same model architecture, training parameters, and dataset usage detailed in Jha \textit{et al.}~\cite{jha2018elemnet}, and the performance of our re-implementation is presented in Table~\ref{table___elemnet_comparison}.

\begin{table}
\centering
\caption{Comparison of MAE score of ElemNet with the original paper.}
\label{table___elemnet_comparison}
\begin{tabular}{c|cc}
\hline
Methods & Original~\cite{jha2018elemnet}& Our replication \\ \hdashline

MAE (eV/atom) & 0.05 &  0.07 \\
\hline
\end{tabular}
\end{table}

\section{SuperDiff and SuperDiff w/ CG} \label{section__apx__superdiff}
We implemented SuperDiff by following the procedures outlined in its original paper and official implementations. Specifically, we divided the dataset used for training the $T_c$ prediction model into cuprates, pnictides, and others, then trained SuperDiff separately for each category. While the $T_c$ prediction model was trained on normalised compositions, SuperDiff was trained without normalisation.

For conditional inference using Iterative Latent Variable Refinement (ILVR), we selected the appropriate SuperDiff model (cuprates, pnictides, or others) based on the reference material. We then applied guidance through “Classifier Guidance (CG)” using both the $T_c$  prediction model and ElemNet with the loss function described in equation~\ref{eq___naive_loss}. Although neither is strictly a classifier, we refer to this approach as CG for convenience. We describe two inference algorithms: one that performs ILVR-based conditional inference with SuperDiff in Algorithm~\ref{alg:superdiff_sampling}, and one that further incorporates the two prediction models as guidance in Algorithm~\ref{alg:superdiff_sampling_w_guidance}. Note that Algorithm~\ref{alg:superdiff_sampling_w_guidance} is not proposed in Yuan \textit{et al.}~\cite{yuan2024diffusion}.

Normally, the classifier used for CG must be trained on data with added noise, which would make Universal Guidance (UG) the better choice for off-the-shelf models. However, our experiments showed that CG still improved $T_c$ without noise-augmented training. Therefore, we decided to use CG. We hypothesize that the reason CG remained effective despite the absence of noise-augmented training is that our $T_c$ prediction model and formation energy prediction model both take normalised compositions as inputs, thereby reducing the impact of noise.

\begin{algorithm}
\caption{SuperDiff using ILVR Sampling Procedures}
\label{alg:superdiff_sampling}
\begin{algorithmic}[1]
%\State \textbf{Input:} Training dataset $\{x_i\}$, number of diffusion steps $T$, noise schedule $\{\beta_t\}_{t=1}^{T}$
%\State \textbf{initialise:} Model parameters $\theta$
%\Procedure{Sampling}{}
  \State $x_T \sim \mathcal{N}(\mathbf{0}, \mathbf{I})$
    %\Comment{initialise from the standard normal distribution}
  \For{$t = T, \dots, 1$}
    \State $z \sim \mathcal{N}(\mathbf{0}, \mathbf{I})$ if \ $t > 1$, else  $z=0$
    \State $x'_{t-1} = \frac{1}{\sqrt{\alpha_t}} \Bigl(x_t - \frac{1 - \alpha_t}{\sqrt{1 - \bar{\alpha}_t}}\,\epsilon_\theta(x_t, t)\Bigr) + \sigma_t z$ \Comment{Denoising}

    \State $x_{t-1} = \phi_N(y_{t-1}) +  x'_{t-1} - \phi_N(x'_{t-1})$
    \Comment{ILVR using sample $y$}
    %\State $ $
  \EndFor
  \State \textbf{return} $x_0$
    \Comment{Generated sample}
%\EndProcedure
\end{algorithmic}
\end{algorithm}

\FloatBarrier

\begin{algorithm}
\caption{SuperDiff using ILVR Sampling Procedures with Classifier Guidance}
\label{alg:superdiff_sampling_w_guidance}
\begin{algorithmic}[1]
%\State \textbf{Input:} Training dataset $\{x_i\}$, number of diffusion steps $T$, noise schedule $\{\beta_t\}_{t=1}^{T}$
%\State \textbf{initialise:} Model parameters $\theta$
%\Procedure{Sampling}{}
  \State $x_T \sim \mathcal{N}(\mathbf{0}, \mathbf{I})$
    %\Comment{initialise from the standard normal distribution}
  \For{$t = T, \dots, 1$}
    \State $z \sim \mathcal{N}(\mathbf{0}, \mathbf{I})$ if \ $t > 1$, else  $z=0$
    \State $x'_{t-1} \leftarrow \frac{1}{\sqrt{\alpha_t}} \Bigl(x_t - \frac{1 - \alpha_t}{\sqrt{1 - \bar{\alpha}_t}}\,\epsilon_\theta(x_t, t)\Bigr) + \sigma_t z$ \Comment{Denoising}
    \State $x_{t-1} \leftarrow x_{t-1} -w \nabla L(x_{t-1})$ \Comment{Guidance with predictors}
    \State $x_{t-1} \leftarrow \phi_N(y_{t-1}) +  x'_{t-1} - \phi_N(x'_{t-1})$ \Comment{ILVR using sample $y$}

  \EndFor
  \State \textbf{return} $x_0$
    \Comment{Generated sample}
%\EndProcedure
\end{algorithmic}
\end{algorithm}

\begin{table}
\centering
\caption{Total changes in $T_c$ resulting from guidance and denoising during the 1000 steps in SuperDiff w/ CG\textbf{ without ILVR}. "Guide weight” denotes the weight $w$ used for guidance in Algorithm~\ref{alg:superdiff_sampling_w_guidance}. "Denoise $\Delta T_c$", "ILVR $\Delta T_c$" and “Guide $\Delta T_c$” represent the cumulative change in $T_c$ per step due to denoising, ILVR, or the guidance. “Sum” is the total of these values. "Screening ratio" denotes the ratio of the number of screened samples to the total number of samples.}
\label{table__result_of_CG_vs_lr_wo_ilvr}
\resizebox{0.5\textwidth}{!}{
\begin{tabular}{ccccc}
\hline

\begin{tabular}{c}Guide \\weight \\$w$\end{tabular} &  \begin{tabular}{c}Denoise\\ $\Delta T_c$(K) \end{tabular} &  \begin{tabular}{c}ILVR\\ $\Delta T_c$(K) \end{tabular} &  \begin{tabular}{c}Guide\\ $\Delta T_c$(K) \end{tabular} &  Sum (K) \\
\hline
    - & 20.0 &     - &- & 20.0 \\
    1.0e-06 & 22.6 &     - &0.1 & 22.7 \\
    1.0e-05 & 20.1 &     - &0.6 & 20.6 \\
    1.0e-04 & 14.0 &     - &5.9 & 19.9 \\
    1.0e-03 &$-$46.5 &     - &     72.7 & 26.2 \\
    1.0e-02 &     $-$491.8 &     - &    545.8 & 54.0 \\
    1.0e-01 &     $-$741.6 &     - &    797.0 & 55.3 \\
    1.0e+00 &$-$10.5 &     - &     33.0 & 22.5 \\
\hline
\end{tabular}
}
\end{table}

\subsection{Guidance vs denoise without ILVR}
We also repeated the experiments from Table~\ref{table__result_of_CG_vs_lr} without using ILVR. As shown in Table~\ref{table__result_of_CG_vs_lr_wo_ilvr}, even without ILVR, excessive increases in $T_c$ are still suppressed by denoising. We hypothesize that properties far from the data distribution are treated as “noise” by the diffusion model and consequently removed.

\section{KIAGO optimisation in the first stage} \label{section__apx__kiago_first_stage}

\begin{figure}
\centering
  \includegraphics[width=\linewidth]{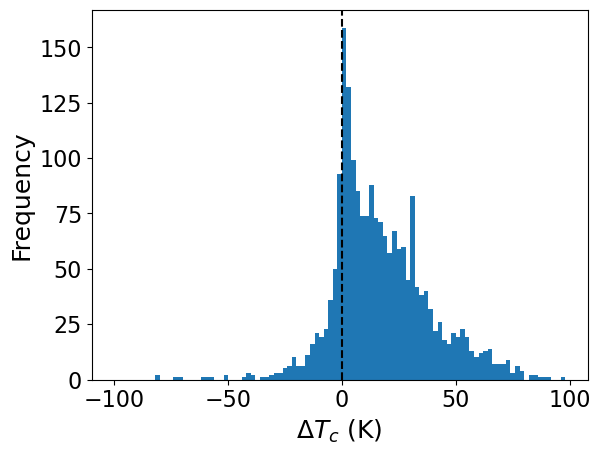}
  \caption{Changes in $T_c$ due to optimisation.}
  \label{figure__tc_optimisation}
\end{figure}

\begin{figure}
\centering
  \includegraphics[width=\linewidth]{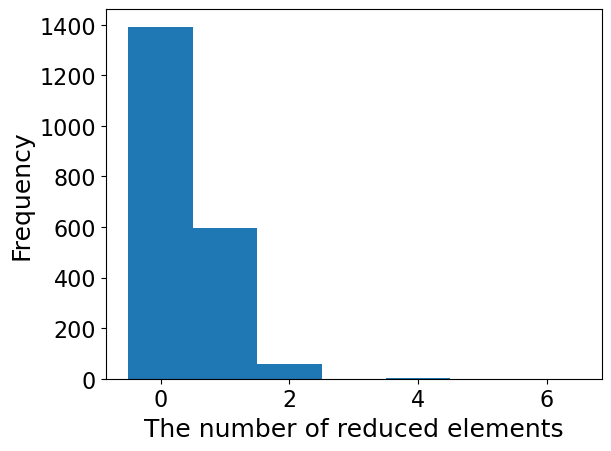}
  \caption{Change in number of element types due to optimisation (a positive number denotes the number of element type decreases after optimisation).}
  \label{figure__element_reduced}
\end{figure}

Here, we show the results of the optimisation using equations~\ref{eq__first_sgate_srt} to \ref{eq__first_sgate_end}. Fig.~\ref{figure__tc_optimisation} shows that most of samples have higher $T_c$ after optimisation. However, some samples have lower $T_c$ after optimisation. This is likely due to the presence of many local minima. Furthermore, we observed a tendency for the number of element types to decrease due to optimisation (Fig.~\ref{figure__element_reduced} ).

%\vspace*{10mm}

\section{Detailed results} \label{section__detailed_results}
Table~\ref{table__result_of_screening_in_exsiting_composition_based} shows the detailed results of the experiments presented in Table~\ref{table__result_of_exsiting_composition_based} in Section~\ref{section__based_on_existing}. Each column in Table~\ref{table__result_of_screening_in_exsiting_composition_based} represents the fraction that fulfills each screening criterion. The last column shows the difference between the average $T_c$ of the top 30 screened samples and the base materials. 

Table~\ref{table__result_of_screening_in_element_substituion} shows the detailed results of the experiments presented in Table~\ref{table__result_of_element_substitution_prob},\ref{table__result_of_element_substitution_neutrality} and~\ref{table__result_of_element_substitution_tc} in Section~\ref{section__element_substitution}. Each column in Table~\ref{table__result_of_screening_in_element_substituion} represents the fraction that fulfills each screening criterion. The last column shows the difference between the average $T_c$ of the top 30 screened samples and the base materials.

%(日本語) 
% (English) section{Detailed results}. Table~\ref{table__result_of_screening_in_exsiting_composition_based} shows the detailed results of the experiments presented in Table~\ref{table__result_of_exsiting_composition_based} in Section~\ref{section__based_on_existing}. Each column in Table~\ref{table__result_of_screening_in_exsiting_composition_based} represents the fraction that fulfills each screening criterion. The last column shows the difference between the average $T_c$ of the top 30 screened samples and the base materials. Table~\ref{table__result_of_screening_in_element_substituion} shows the detailed results of the experiments presented in Table~\ref{table__result_of_element_substitution} in Section~\ref{section__element_substitution}. Each column in Table~\ref{table__result_of_element_substitution} represents the fraction that fulfills each screening criterion. The last column shows the difference between the average $T_c$ of the top 30 screened samples and the base materials.

\begin{table*}
\centering \small
\caption{All results of Screening rates and $T_c$. “Unique rate” indicates the fraction of unique compositions among the 4,096 samples. “ELM < 10,” “Neutrality,” “ElecNeg,” and “$E_f$ < 0.0 eV” denote, respectively, the percentage of these unique samples that: contain at most 9 elements (“ELM < 10”); satisfy charge neutrality per SMACT (“Neutrality”); meet electronegativity criterion using SMACT (“ElecNeg”); achieve predicted formation energy below 0.0 eV ('$E_f$ < 0.0 eV'). “Screened rate” is the fraction that fulfills all these screening criteria among all 4,096 samples. “$\Delta T_c$ Top-30 (K)” represents the difference between the average predicted $T_c$ of the top 30 screened samples and the base materials.}
\label{table__result_of_screening_in_exsiting_composition_based}]
\resizebox{\textwidth}{!}{
\begin{tabular}{l:l|wc{1.3cm}wc{1.3cm}wc{1.3cm}wc{1.3cm}wc{1.3cm}:cc}
\hline
    \begin{tabular}{c}Base materials \\ from SuperCon\end{tabular} & Method &  Unique rate &  ELM < 10 &  Neutrality &  ElecNeg &  $E_f$< 0.0 eV &  Screened rate &  $\Delta T_c$ Top-30 (K) \\
    \hline
    \multirow{4}{*}{$\mathrm{LaNiAsO}$} & KIAGO & 0.94 & 1.00 & 0.73 & 0.58 & 0.98 & 0.57 & 97.64 \\  
     & SD & 1.00 & 0.50 & 0.08 & 0.01 & 0.96 & 0.01 & $-$1.42 \\  
     & SD w/ CG & 1.00 & 0.49 & 0.08 & 0.01 & 0.97 & 0.01 & $-$0.48 \\  
     & C-ES & 0.89 & 1.00 & 0.99 & 0.84 & 1.00 & 0.84 & 16.79 \\ 
    \hline
    \multirow{4}{*}{$\mathrm{SrFe_{1.88}Ni_{0.12}As_{2}}$} & KIAGO & 0.99 & 1.00 & 0.79 & 0.59 & 0.64 & 0.37 & 86.77 \\  
     & SD & 1.00 & 0.46 & 0.05 & 0.00 & 0.21 & 0.00 & 26.76 \\  
     & SD w/ CG & 1.00 & 0.49 & 0.04 & 0.01 & 0.23 & 0.00 & 0.74 \\  
     & C-ES & 0.48 & 1.00 & 0.98 & 0.72 & 0.24 & 0.19 & 18.96 \\ 
    \hline
    \multirow{4}{*}{$\mathrm{Sr_{4}V_{2}Fe_{2}As_{2}O_{6}}$} & KIAGO & 0.91 & 1.00 & 0.70 & 0.60 & 0.91 & 0.55 & 89.89 \\  
     & SD & 1.00 & 0.12 & 0.29 & 0.05 & 0.51 & 0.01 & $-$13.77 \\  
     & SD w/ CG & 1.00 & 0.02 & 0.25 & 0.02 & 0.61 & 0.00 & $-$3.99 \\  
     & C-ES & 0.85 & 1.00 & 1.00 & 0.96 & 1.00 & 0.96 & $-$0.00 \\ 
    \hline
    \multirow{4}{*}{$\mathrm{LaPt_{2}B_{2}C}$} & KIAGO & 0.95 & 1.00 & 0.71 & 0.51 & 0.96 & 0.49 & 86.39 \\  
     & SD & 1.00 & 0.52 & 0.06 & 0.01 & 0.87 & 0.01 & $-$5.01 \\  
     & SD w/ CG & 1.00 & 0.38 & 0.08 & 0.02 & 0.90 & 0.01 & $-$5.27 \\  
     & C-ES & 0.89 & 1.00 & 0.75 & 0.53 & 0.65 & 0.22 & 5.06 \\ 
    \hline
    \multirow{4}{*}{$\mathrm{HgBa_{2}Ca_{2}Cu_{3}O_{8}}$} & KIAGO & 0.74 & 1.00 & 0.65 & 0.55 & 0.96 & 0.54 & 16.70 \\  
     & SD & 1.00 & 0.57 & 0.00 & 0.00 & 1.00 & 0.00 & $-$29.43 \\  
     & SD w/ CG & 1.00 & 0.58 & 0.01 & 0.00 & 1.00 & 0.00 & $-$21.36 \\  
     & C-ES & 0.87 & 1.00 & 0.99 & 0.93 & 1.00 & 0.93 & 1.00 \\ 
    \hline
    \multirow{4}{*}{$\mathrm{CeBiS_{2}O}$} & KIAGO & 0.81 & 1.00 & 0.63 & 0.53 & 0.88 & 0.47 & 94.89 \\  
     & SD & 1.00 & 0.21 & 0.14 & 0.01 & 0.23 & 0.00 & N/A \\  
     & SD w/ CG & 1.00 & 0.05 & 0.31 & 0.08 & 0.22 & 0.00 & N/A \\  
     & C-ES & 0.85 & 1.00 & 1.00 & 0.94 & 0.49 & 0.48 & 0.42 \\ 
    \hline
    \multirow{4}{*}{$\mathrm{Bi_{2}Sr_{2}CuO_{6}}$} & KIAGO & 0.81 & 1.00 & 0.68 & 0.57 & 0.95 & 0.55 & 127.33 \\  
     & SD & 1.00 & 0.76 & 0.01 & 0.01 & 1.00 & 0.01 & 18.34 \\  
     & SD w/ CG & 1.00 & 0.76 & 0.01 & 0.00 & 1.00 & 0.00 & 14.38 \\  
     & C-ES & 0.88 & 1.00 & 0.99 & 0.94 & 1.00 & 0.94 & 52.25 \\ 
    \hline
    \multirow{4}{*}{$\mathrm{TlSr_{2}CaCu_{2}O_{7}}$} & KIAGO & 0.74 & 1.00 & 0.67 & 0.57 & 0.94 & 0.55 & 74.64 \\  
     & SD & 1.00 & 0.67 & 0.01 & 0.01 & 1.00 & 0.00 & 5.21 \\  
     & SD w/ CG & 1.00 & 0.70 & 0.01 & 0.00 & 1.00 & 0.00 & 11.53 \\  
     & C-ES & 0.87 & 1.00 & 0.99 & 0.94 & 1.00 & 0.94 & 34.07 \\ 
    \hline
\end{tabular}
}
\end{table*}

\begin{table*}
\centering \small
\caption{“Keeping remaining atoms” is the percentage of generated compositions in which the designated set of elements and their compositions are preserved within a 1\% error (e.g., In the top row of the table, the composition of $\mathrm{FeAsF_{0.2}O_{0.8}}$ in $\mathrm{CeFeAsF_{0.2}O_{0.8}}$.) “Keeping sum of substitute atoms” is the percentage in which the total substituted composition—excluding the designated elements and compositions—matches the targeted substituted element's composition within 1\% error (e.g., In the top row of the table, the composition of $\mathrm{Ce^{+3}}$ in $\mathrm{CeFeAsF_{0.2}O_{0.8}}$.). Other metrics follow the definitions in Table~\ref{table__result_of_screening_in_exsiting_composition_based}.}
\label{table__result_of_screening_in_element_substituion}
\resizebox{\textwidth}{!}{
\begin{tabular}{wc{1.5cm}wc{1.0cm}wc{1.0cm}|wc{1.0cm}wc{1.0cm}wc{1.0cm}wc{1.0cm}wc{1.0cm}wc{1.0cm}wc{1.0cm}:wc{1.0cm}wc{1.0cm}}
\hline
    \begin{tabular}{c}Base materials \\ from SuperCon\end{tabular} &  \begin{tabular}{c}Substitute\\target\end{tabular} & Method &  \begin{tabular}{c}Unique\\rate\end{tabular} &  ELM < 10 &  Neutrality &  ElecNeg &  $E_f$< 0.0 eV &  \begin{tabular}{c}Keeping\\sum of\\substitute\\atoms\end{tabular} &  \begin{tabular}{c}Keeping\\remain\\atoms\end{tabular} &  \begin{tabular}{c}Screening\\rate\end{tabular} &  \begin{tabular}{c}$\Delta T_c$ \\Top-30 (K)\end{tabular} \\
    \hline
    \multirow{4}{*}{$\mathrm{CeFeAsF_{0.2}O_{0.8}}$} & \multirow{4}{*}{$\mathrm{Ce^{+3}}$} & KIAGO & 0.99 & 1.00 & 1.00 & 0.26 & 0.99 & 1.00 & 1.00 & 0.25 & 14.53 \\  
     & & SD & 1.00 & 0.40 & 0.03 & 0.01 & 0.54 & 0.00 & 0.13 & 0.00 & N/A \\  
     & & SD w/ CG & 1.00 & 0.34 & 0.03 & 0.01 & 0.47 & 0.00 & 0.13 & 0.00 & N/A \\  
     & & C-ES & 0.81 & 1.00 & 1.00 & 0.45 & 0.94 & 1.00 & 1.00 & 0.42 & 9.99 \\ 
    \hline
    \multirow{4}{*}{$\mathrm{LaFeAsO_{1.0}}$} & \multirow{4}{*}{$\mathrm{La^{+3}}$} & KIAGO & 1.00 & 1.00 & 1.00 & 0.09 & 1.00 & 1.00 & 1.00 & 0.09 & 35.62 \\  
     & & SD & 1.00 & 0.45 & 0.02 & 0.01 & 0.58 & 0.04 & 0.18 & 0.00 & 1.24 \\  
     & & SD w/ CG & 1.00 & 0.39 & 0.01 & 0.00 & 0.51 & 0.04 & 0.18 & 0.00 & N/A \\  
     & & C-ES & 0.85 & 1.00 & 1.00 & 0.41 & 0.98 & 1.00 & 1.00 & 0.39 & 29.45 \\ 
    \hline
    \multirow{4}{*}{$\mathrm{SrFe_{2}As_{2}}$} & \multirow{4}{*}{$\mathrm{Sr^{+2}}$} & KIAGO & 1.00 & 1.00 & 1.00 & 0.17 & 0.26 & 1.00 & 1.00 & 0.06 & 29.35 \\  
     & & SD & 1.00 & 0.52 & 0.03 & 0.00 & 0.21 & 0.04 & 0.15 & 0.00 & N/A \\  
     & & SD w/ CG & 1.00 & 0.44 & 0.07 & 0.01 & 0.20 & 0.05 & 0.13 & 0.00 & N/A \\  
     & & C-ES & 0.87 & 1.00 & 1.00 & 0.34 & 0.23 & 1.00 & 1.00 & 0.09 & 13.78 \\ 
    \hline
    \multirow{4}{*}{$\mathrm{Bi_{2}CaSr_{2}Cu_{2}O_{8}}$} & \multirow{4}{*}{$\mathrm{Bi^{+3}}$} & KIAGO & 0.96 & 1.00 & 1.00 & 1.00 & 1.00 & 1.00 & 1.00 & 1.00 & 14.61 \\  
     & & SD & 1.00 & 0.70 & 0.01 & 0.01 & 1.00 & 0.41 & 0.56 & 0.00 & 0.15 \\  
     & & SD w/ CG & 1.00 & 0.65 & 0.01 & 0.01 & 1.00 & 0.39 & 0.52 & 0.00 & 0.42 \\  
     & & C-ES & 0.82 & 1.00 & 1.00 & 0.94 & 1.00 & 1.00 & 1.00 & 0.94 & 6.42 \\ 
    \hline
    \multirow{4}{*}{$\mathrm{CeNiC_{2}}$} & \multirow{4}{*}{$\mathrm{Ce^{+4}}$} & KIAGO & 0.99 & 1.00 & 1.00 & 0.40 & 1.00 & 1.00 & 1.00 & 0.40 & 17.30 \\  
     & & SD & 1.00 & 0.72 & 0.07 & 0.03 & 0.98 & 0.02 & 0.04 & 0.00 & N/A \\  
     & & SD w/ CG & 1.00 & 0.77 & 0.06 & 0.02 & 0.98 & 0.02 & 0.04 & 0.00 & N/A \\  
     & & C-ES & 0.88 & 1.00 & 1.00 & 0.55 & 0.93 & 1.00 & 1.00 & 0.50 & 6.26 \\ 
    \hline
    \multirow{4}{*}{$\mathrm{LaNiC_{2}}$} & \multirow{4}{*}{$\mathrm{La^{+3}}$} & KIAGO & 1.00 & 1.00 & 1.00 & 0.82 & 1.00 & 1.00 & 1.00 & 0.82 & 25.24 \\  
     & & SD & 1.00 & 0.76 & 0.03 & 0.01 & 0.98 & 0.07 & 0.12 & 0.00 & N/A \\  
     & & SD w/ CG & 1.00 & 0.76 & 0.04 & 0.01 & 0.98 & 0.07 & 0.11 & 0.00 & 0.12 \\  
     & & C-ES & 0.88 & 1.00 & 1.00 & 0.70 & 0.93 & 1.00 & 1.00 & 0.64 & 7.55 \\ 
    \hline
    \multirow{4}{*}{$\mathrm{MgCoNi_{3}}$} & \multirow{4}{*}{$\mathrm{Co^{+2}}$} & KIAGO & 1.00 & 1.00 & 1.00 & 0.96 & 0.00 & 1.00 & 1.00 & 0.00 & N/A \\  
     & & SD & 1.00 & 0.42 & 0.71 & 0.65 & 0.36 & 0.01 & 0.02 & 0.00 & N/A \\  
     & & SD w/ CG & 1.00 & 0.42 & 0.64 & 0.59 & 0.38 & 0.01 & 0.02 & 0.00 & N/A \\  
     & & C-ES & 0.88 & 1.00 & 1.00 & 0.35 & 0.24 & 1.00 & 1.00 & 0.07 & $-$2.41 \\ 
    \hline
    \multirow{4}{*}{$\mathrm{RuSr_{2}GdCu_{2}O_{8}}$} & \multirow{4}{*}{$\mathrm{Sr^{+2}}$} & KIAGO & 0.95 & 1.00 & 1.00 & 1.00 & 1.00 & 1.00 & 1.00 & 1.00 & $-$33.66 \\  
     & & SD & 1.00 & 0.55 & 0.05 & 0.04 & 1.00 & 0.13 & 0.21 & 0.00 & $-$0.23 \\  
     & & SD w/ CG & 1.00 & 0.57 & 0.05 & 0.04 & 1.00 & 0.14 & 0.22 & 0.00 & $-$0.08 \\  
     & & C-ES & 0.81 & 1.00 & 1.00 & 0.90 & 1.00 & 1.00 & 1.00 & 0.90 & 3.70 \\ 
    \hline
    \multirow{4}{*}{$\mathrm{RuSr_{2}YCu_{2}O_{8}}$} & \multirow{4}{*}{$\mathrm{Y^{+3}}$} & KIAGO & 0.85 & 1.00 & 1.00 & 1.00 & 1.00 & 1.00 & 1.00 & 1.00 & 46.16 \\  
     & & SD & 1.00 & 0.65 & 0.08 & 0.06 & 1.00 & 0.09 & 0.29 & 0.00 & $-$0.89 \\  
     & & SD w/ CG & 1.00 & 0.64 & 0.07 & 0.05 & 1.00 & 0.12 & 0.28 & 0.00 & 2.02 \\  
     & & C-ES & 0.81 & 1.00 & 1.00 & 0.94 & 1.00 & 1.00 & 1.00 & 0.94 & 37.60 \\ 
    \hline
    \multirow{4}{*}{$\mathrm{Y_{2}Fe_{3}Si_{5}}$} & \multirow{4}{*}{$\mathrm{Y^{+3}}$} & KIAGO & 1.00 & 1.00 & 1.00 & 0.67 & 1.00 & 1.00 & 1.00 & 0.67 & 7.56 \\  
     & & SD & 1.00 & 0.33 & 0.18 & 0.05 & 0.56 & 0.03 & 0.05 & 0.00 & N/A \\  
     & & SD w/ CG & 1.00 & 0.31 & 0.21 & 0.07 & 0.58 & 0.03 & 0.04 & 0.00 & N/A \\  
     & & C-ES & 0.87 & 1.00 & 1.00 & 0.16 & 1.00 & 1.00 & 1.00 & 0.16 & 1.45 \\ 
    \hline
    \multirow{4}{*}{$\mathrm{YIrSi}$} & \multirow{4}{*}{$\mathrm{Y^{+3}}$} & KIAGO & 1.00 & 1.00 & 1.00 & 0.02 & 1.00 & 1.00 & 1.00 & 0.02 & 5.08 \\  
     & & SD & 1.00 & 0.90 & 0.28 & 0.06 & 0.70 & 0.02 & 0.04 & 0.00 & 3.42 \\  
     & & SD w/ CG & 1.00 & 0.90 & 0.30 & 0.07 & 0.72 & 0.02 & 0.03 & 0.00 & $-$1.43 \\  
     & & C-ES & 0.87 & 1.00 & 1.00 & 0.34 & 0.99 & 1.00 & 1.00 & 0.34 & 5.31 \\ 
    \hline
\end{tabular}
}
\end{table*}

\end{document}